\documentclass[10pt,nonatbib,reprint]{common/sigplanconf}
\usepackage{amsmath}
\usepackage{times}
\usepackage{xspace}
\usepackage{datetime}
\usepackage[hyphens]{url}
\usepackage{hyperref}
\usepackage[usenames,dvipsnames]{color}
\usepackage{graphicx}
\usepackage{epsfig}
\usepackage[table,xcdraw]{xcolor}
\usepackage{multirow}
\graphicspath{{figures/}}
\usepackage{amssymb}
\usepackage{algorithm}
\usepackage[noend]{algpseudocode}
\usepackage{cite}
\usepackage{caption}
\usepackage{pgfplots}
\conferenceinfo{SOSP'17}{October 29--31, 2017, Shanghai, China}
\copyrightyear{2017}

\newcommand{\coloredcomment}[3]{}

\newcommand{\ulfar}[1]{\coloredcomment{ulfar}{#1}{cyan}}

\newcommand{\marginulfar}[1]{\marginpar{\ulfar{#1}}}
\newcommand{\deadcomment}[1]{}
\usepackage{textgreek}
\usepackage{mathtools}
\newcommand{\ochlo}{{{\rlap{$\acute{\;\,}$}\textomikron}\textchi\textlambda\textomikron\textvarsigma}\xspace}

\newcommand{\prochlo}{\textsc{Prochlo}\xspace}
\newcommand{\ESA}{\textsc{ES\kern -1ptA}\xspace}

\preprintfooter{Draft shepherding revision of {\currenttime}, \today{}}

\date{}

\authorinfo{Andrea Bittau$^\star$ %
   $\;\;$  {{\'{U}}}lfar Erlingsson$^\star$ %
   $\;\;$  Petros Maniatis$^\star$ %
   $\;\;$  Ilya Mironov$^\star$ %
   $\;\;$  Ananth Raghunathan$^\star$ %
   $\;\;$  David Lie$^\ddagger$ %
   $\;\;$  Mitch Rudominer$^\circ$ %
   $\;\;$  Ushasree Kode$^\circ$ %
   $\;\;$  Julien Tinnes$^\circ$ %
   $\;\;$  Bernhard Seefeld$^\circ$%
}{
  $^\star$Google Brain                   \hspace*{3em} 
  $^\ddagger$Google Brain and U.\ Toronto  \hspace*{3em} 
  $^\circ$Google
}

\begin{document}


\CopyrightYear{2017}
\setcopyright{rightsretained}
\conferenceinfo{SOSP '17,}{October 28, 2017, Shanghai, China}
\isbn{978-1-4503-5085-3/17/10}
\acmPrice{}
\doi{https://doi.org/10.1145/3132747.3132769}
\title{\prochlo: Strong Privacy for Analytics in the Crowd}
\maketitle



\begin{abstract}
  The large-scale monitoring
  of computer users' software activities
  has become commonplace,
  e.g., for application telemetry, error reporting, or demographic profiling.
  %
  %
  This paper describes a principled systems architecture---\emph{Encode, Shuffle, Analyze} (\ESA)---for
  performing such monitoring with high utility while also protecting user privacy.
  The \ESA design, and its \prochlo implementation,
  are informed by our practical experiences
  with an existing, large deployment of privacy-preserving software monitoring.

  With \ESA, the privacy of monitored users' data is guaranteed 
  by its processing
  in a three-step pipeline.
  First, the data is \emph{encoded} to control scope, granularity, and randomness.
  Second, the encoded data is collected in batches subject to a randomized threshold,
  and blindly \emph{shuffled}, to break linkability and
  to ensure that individual data items get ``lost in the crowd'' of the batch.
  Third,
  the anonymous, shuffled data\marginulfar{too many hypens}
  is \emph{analyzed} by a specific analysis engine that further
  prevents statistical inference attacks on analysis results.

  \ESA 
  extends existing best-practice methods for sensitive-data analytics,
  by using cryptography and statistical techniques
  to make explicit 
  how data is elided and reduced in precision,\marginulfar{long sentence. break up? example?}
  how only common-enough, anonymous data is analyzed,
  and how this is done for only specific, permitted purposes.
  As a result,
  \ESA remains compatible with 
  the established workflows
  of traditional database analysis.

  Strong privacy guarantees, including differential privacy, can be established
  at each processing step to defend against malice or compromise at one or more
  of those steps.  \prochlo develops new techniques to harden 
  those steps, including the \emph{Stash Shuffle}, a novel scalable and
  efficient oblivious-shuffling algorithm based on Intel's SGX,
  and new applications of cryptographic secret sharing and blinding.
  We describe \ESA and \prochlo, 
  as well as experiments that validate their ability to balance utility and privacy.
\end{abstract}

\section{Introduction}
\label{sec:intro}
Online monitoring of client software behavior
has long been used 
for disparate purposes,
such as measuring feature adoption
or performance characteristics,
as well as large-scale error-reporting~\cite{glerum2009debugging}.
For modern software,
such monitoring
may entail
systematic collection of information about client devices, their users, and the software they run~\cite{cloud_apps,web_apps,mobile_apps}.
This data collection is
in many ways fundamental to
modern software operations and economics,
and provides many clear benefits, 
e.g., it enables the deployment of security updates
that eliminate software vulnerabilities~\cite{chrome_security_lessons}.

For such data,
the processes, mechanisms, and other means
of privacy protection
are an increasingly high-profile concern.
This is especially true
when data is
collected automatically
and when it is utilized
for building user profiles or demographics~\cite{freeforall,mobile_apps,tracking_trackers}.
Regrettably,
in practice,
those concerns 
often remain unaddressed,
sometimes despite the existence of strong incentives that would suggest otherwise.
One reason for this
is that
techniques that can guarantee privacy
exist mostly as theory,
as limited-scope deployments,
or as innovative-but-nascent mechanisms~\cite{dwork2008differential,rappor,guha_new,abadi_csf17}.

We introduce the \emph{Encode, Shuffle, Analyze} (\ESA) architecture
for privacy-preserving software monitoring,
and its \prochlo implementation.\footnote{\prochlo combines privacy with the Greek word {\ochlo} for crowds.}
The \ESA architecture
is informed by our experience
building, operating, and maintaining
the RAPPOR privacy-preserving monitoring system
for the Chrome Web browser~\cite{rappor}.
Over the last 3 years, RAPPOR has processed
up to billions of daily, randomized reports
in a manner that guarantees local differential privacy,
without assumptions about users' trust;
similar techniques
have since gained
increased 
attention~\cite{apple_korolova,somesh_usenix,guha_new,livshits_usenix}.
However,
these techniques
have limited utility,
both in theory and in our experience,
and their
statistical nature
makes them ill-suited
to standard software engineering practice.

Our \ESA architecture 
overcomes the limitations of systems like RAPPOR,
by extending and strengthening
current best practices in private-data processing.
In particular,
\ESA
enables any high-utility analysis algorithm
to be compatible with strong privacy guarantees,
by appropriately building on users' trust assumptions,
privacy-preserving randomization,
and cryptographic mechanisms---in
a principled manner, encompassing realistic attack models.

With \ESA,
software users' data
is collected into a database of encrypted records
to be processed
only by a specific \emph{analysis},
determined by the corresponding data decryption key.
Being centralized,
this database is compatible with existing,
standard software engineering processes,
and its analysis may use
the many known techniques
for balancing utility and privacy,
including differentially-private release~\cite{DworkRoth}.

The contents of a materialized
\ESA database 
are guaranteed to have been \emph{shuffled}
by a party that is trusted
to remove implicit and indirect identifiers,
such as arrival time, order, or originating IP address.
Shuffling
operates on batches of data
which are collected over an \emph{epoch} of time
and reach a minimum threshold cardinality.
Thereby, shuffling ensures
that each record becomes part of a \emph{crowd}---not
just by hiding timing and metadata, 
but also by ensuring that more than 
a (randomized) threshold number of reported items exists for every different data partition
forwarded for analysis.

Finally,
each \ESA database record
is constructed
by specific reporting software at a user's client.
This software ensures that records are
\emph{encoded} for privacy,
e.g., by removing direct identifiers or extra data,
by fragmenting 
to make the data less unique,
or by adding noise for local differential privacy or plausible deniability.
Encoded data is transmitted 
using nested encryption
to ensure that the data will be shuffled and analyzed only by the shuffler and analyzer in possession of the corresponding private keys.

\ESA protects privacy
by building on users'
existing trust relationships
as well as technical mechanisms.
Users indicate their trust assumptions
by installing and executing client-side software
with embedded cryptographic keys for a specific shuffler and analyzer.
Based on those keys,
the software's
reporting features encode and transmit data
for exclusive processing by the subsequent steps.
Each of these processing steps
may independently take measures to protect privacy.
In particular,
by adding random noise to data, thresholds, or analysis results,
each step may guarantee
differential privacy (as discussed in detail in \S\ref{sec:dp_guarantees}).
\smallskip


\noindent Our contributions are as follows:\vspace*{-1ex}
\begin{itemize}
\item We describe the \ESA architecture, which enables
  high-utility, large-scale monitoring of users' data
  while protecting their privacy.
  Its novelty stems from
  explicit trust assumptions captured by nested encryption,
  guarantees of anonymity and uncertainty added by a shuffler intermediary,
  and a pipeline where
  differential privacy guarantees can be independently added at each stage.
  
\item We describe \prochlo, a hardened implementation of \ESA that uses
  SGX~\cite{sgx}, showing strong privacy guarantees to be compatible with
  efficiency, scalability, and a small trusted computing base. We design,
  implement, and evaluate a new oblivious-shuffling algorithm for SGX, the
  \emph{Stash Shuffle}, that scales efficiently to our target data sizes,
  unlike previous algorithms.  We also introduce new cryptographic
  primitives for secret-shared, blind data thresholding, reducing
  users' trust requirements.
  
\item We evaluate \prochlo on two representative monitoring use cases,
  as well as a collaborative filtering task and a deep-learning task,
  in each case
  achieving both high-utility
  and strong privacy guarantees
  in their analysis.
\end{itemize}
On this basis
we conclude that 
the \ESA architecture for monitoring users' data
significantly advances the practical state-of-the-art,
compared to previous, deployed systems.

\section{Motivation and Alternatives}
\label{sec:motivation}
The systems community 
has long recognized
the privacy concerns raised by software monitoring,
and addressed them in various ways,
e.g., by authorization and access controls.
Notably,
for large-scale error reporting and profiling,
it has become established best practice 
for systems software to
perform data reduction, scrubbing, and minimization---respectively,
to eliminate superfluous data,
remove unique identifiers,
and to coarsen and minimize what is collected
without eliminating statistical signals---and to require
users' opt-in approval for each collected report.
For example,
large-scale Windows error reporting 
operates in this manner~\cite{glerum2009debugging}.

However,
this established practice
is ill-suited to automated monitoring at scale,
since it is based solely on 
users' declared trust in the collecting party
and explicit approval of each transmitted report,
without any technical privacy guarantees.
Perhaps as a result,
there has been little practical deployment
of the many promising mechanisms
based on pervasive monitoring
developed by the systems community,
e.g., for understanding
software behavior or bugs,
software performance or resource use,
or for other software
improvements~\cite{swe_needs,performance_bugs,statistical_bugs,battery_bugs,appinsight,mobile_tracing}.
This is ironic,
because these mechanisms
rely on common statistics and correlations---not
individual users' particular data---and
should therefore be compatible with protecting users' privacy.

To serve these purposes, the \ESA architecture
is a flexible platform 
that allows
high-utility analysis of
software-monitoring data,
without increasing the privacy risk of users.
For ease of use,
\ESA 
is compatible with, and extends,
existing software engineering processes
for sensitive-data analytics.
These include
data elimination, coarsening, scrubbing, and anonymization,
both during collection and storage,
as well as careful access controls during analysis,
and public release of only privacy-preserving data~\cite{rappor}.

\subsection{A Systems Use Case for Software Monitoring}
\label{sec:mot:example}
As a concrete motivating example
from the domain of systems software,
consider the task of
determining 
which system APIs
are used by which individual software application.
This task
highlights
the clear benefits of large-scale software monitoring,
as well as the difficulty of protecting users' privacy.
Just one of its many potential benefits
is the detection of seldom-used APIs
and the cataloging of applications still using old, legacy APIs,
such that those APIs can be deprecated and removed in future system versions.
\ulfar{(Other uses include monitoring API adoption,
  detecting anomalies in API usage, etc.}

Na\"ively collecting monitoring data for this task 
may greatly affect users' privacy,
since
both the identity of applications and their API usage profiles
are closely correlated to user activities---including
those illegal, embarrassing, or otherwise damaging,
such as copyright infringement, gambling, etc.
Indeed,\marginulfar{Do we need example here?}
both applications and the combinations of APIs they use
may be unique, incriminating, or secret (e.g., for developers of their own applications).
Therefore,
to guarantee privacy,
this task must be realized
without associating users with the monitored data,
without revealing secret applications or API usage patterns,
and without creating a database of sensitive (or deanonymizable) data---since
such databases
are at risk for abuse, compromise, or theft.

This paper describes
how 
this monitoring task
can be achieved---with privacy---using
the \ESA architecture.
However,
we must first
consider
what is meant by privacy,
and
the alternative means of guaranteeing privacy.


\subsection{Privacy Approaches, Experience, and Refinements}
\label{sec:mot:lessons}
Generally, privacy
is intuitively understood
as
individuals' socially-defined ability
to control the release of 
their personal information~\cite{abadi_csf17,saltzer_schroeder}.
As such, privacy is not easily characterized
in terms of traditional computer security properties.
For example, despite the integrity and secrecy of input data about individuals,
social perceptions
about those individuals
may still be
changed if
the observable output of computations
enables new statistical inferences~\cite{denning80s}.
\smallskip

\noindent\textit{\textbf{Privacy Approaches and Differential Privacy Definitions}}\\
Myriad new definitions and techniques have been proposed
for privacy protection---mostly
without much traction or success.
For example,
the well-known property of \emph{$k$-anonymity}
prevents direct release of information about
subgroups with fewer than $k$ individuals~\cite{kanon,Samarati:2001:PRI:627337.628183}.
However,
since
$k$-anonymity
does not limit
statistical inferences
drawn indirectly,
from subgroup intersections or differences,
it cannot
offer mathematically-strong privacy guarantees~\cite{k-anonymity-attacks,machanavajjhala2006diversity}.

In the last decade,
the definitions of
differential privacy
have become the accepted standard for strong privacy~\cite{DMNS,DworkRoth}.
The guarantees of
differential privacy
stem from the uncertainty induced 
by adding
carefully-selected random noise
to individuals' data,
to the intermediate values computed from that data,
or to the final output of analysis computations.
As a result,
each-and-every individual
may be sure
that \emph{all} statistical inferences
are insensitive to their data;
that is, 
their participation in an analysis
makes no difference,
to their privacy loss---even in the worst case,
for the unluckiest individual.
Specifically,
the guarantees 
establish an $\varepsilon$ upper bound 
on the sensitivity in at least $1-\delta$ cases;
thus,
an $(\varepsilon,\delta)$-private epidemiology study of smoking and cancer
can guarantee
(except, perhaps, with $\delta$ probability)
that attackers' guesses about each participant's
smoking habits or illnesses
can change by a multiplicative $e^\varepsilon$ factor, at most.

A wide range of mechanisms
have been developed for the
\emph{differentially-private release} 
of analysis results.
Those mechanisms support
analysis ranging from simple statistics
through domain-specific network analysis,
to general-purpose deep learning of neural networks
by stochastic gradient descent~\cite{DworkRoth,FrankSigcomm,abadi2016deep}.
Alas,
most of these 
differentially-private mechanisms
assume that
analysis is performed on
a trusted, centralized database platform
that can
securely protect
the long-term secrecy and integrity
of both individuals' data
and the analysis itself.
This is disconcerting,
because it entails that any platform compromise
can arbitrarily impact privacy---e.g., by making individuals' data public---and
the risk of compromise seems real,
since
database breaches are all too common~\cite{us-office-of-personnel-management-breach,equifax-breach}.

Instead of
adding uncertainty to analysis output
to achieve differential privacy,
an alternative is to
add random noise
to each individual's data, before it is collected.
This can be done
using techniques that provide local differential privacy
(e.g., randomized response)
in a manner that permits moderately good utility~\cite{rappor,apple_korolova,somesh_usenix,guha_new}.
By adopting this approach,
the risk of database compromise can be fully addressed:
the data collected from each individual
has built-in privacy guarantees,
which hold without further assumptions
and require no trust in other parties.
\smallskip

\noindent\textit{\textbf{Experiences with Local Differential Privacy Mechanisms}}\\
Local differential privacy
is the approach taken
by our RAPPOR software monitoring system
for the Chrome Web browser~\cite{rappor}.
Over the last 3 years, we have 
deployed, operated, and maintained
RAPPOR versions 
utilized
for hundreds of disparate purposes,
by dozens of software engineers,
to process
billions of daily, randomized reports~\cite{rappor-chrome}.

Regrettably,
there are strict limits
to the utility
of locally-differentially-private analyses.
Because each reporting individual
performs independent coin flips,
any analysis results
are perturbed by noise
induced by the properties of the binomial distribution.
The magnitude of this random Gaussian noise can be very large:
even in the theoretical best case,
its standard deviation 
grows in proportion to the square root of the report count,
and the noise
is in practice higher by an order of magnitude~\cite{rappor,somesh_usenix,guha_new,unknown_rappor,evfimievski2003limiting}.
Thus,
if a billion individuals' reports are analyzed,
then a common signal from even up to a million reports
may be missed.

Because of their inherent noise,
local differential privacy approaches
are best suited for
measuring the most frequent elements
in data from peaky power-law distributions.
This greatly limits their applicability---although they
may sometimes be a good fit, 
such as for
RAPPOR's initial purpose of  tracking common, incorrect software configurations.
For example,
a thousand apps and a hundred APIs
may be relevant to
a task of measuring applications' API usage (\S\ref{sec:mot:example}).
However,
an infeasible amount of data
is required to get
a clear signal for each app/API combination
with local differential privacy:
an order-of-magnitude more than 100,000, squared, i.e.,
reports from one trillion individual users~\cite{unknown_rappor}.

Somewhat surprisingly,
if the set of reports can be partitioned in the right manner,
the utility of RAPPOR (and similar systems)
can be greatly enhanced
by analyzing fewer reports at once.
By placing correlated data into the same partitions,
signal recovery
can be facilitated,
especially since
the square-root-based noise floor
will be lower in each partition
than in the entire dataset.

In particular,
the data required for the app/API example above
can be reduced by
two orders of magnitude,
if the reported API is used to partition RAPPOR reports
into 100 disjoint, separately-analyzed sets;
for each separate API,
only 100 million reports
are required to find the top 1000 apps,
by the above arithmetic
(see also \S\ref{sec:eval:vocab}'s  experiment).

Unfortunately,
such partitioning may greatly
weaken privacy guarantees:
differential privacy
is fundamentally incompatible
with the certain knowledge that an API was used,
let alone that a particular individual used that API~\cite{denning80s,k-anonymity-attacks}.
Therefore,
any such partitioning must be done with great care
and in a way that adds uncertainty
about each partition.

Another major obstacle
to the practical use of 
locally-differentially-private methods---based on our experiences with RAPPOR---is
the opaque, fixed, and statistical nature of the data collected.
Not only does this
prevent exploratory data analysis
and any form of manual vetting,
but it also
renders the reported data incompatible with
the existing tools and processes
of standard engineering practice.
Even when some users (e.g., of Beta or Developer software versions)
have opted into reporting
more complete data,
this data 
is not easily correlated with other reports
because it is not collected via the same pipelines.
In our experience,
this is a frustrating obstacle to developers,
who have been unable to
get useful signals from RAPPOR analysis
in a substantial fraction of cases,
due to
the noise added for privacy,
and the difficulties of 
setting up monitoring and interpreting results.
\smallskip

\noindent\textit{\textbf{Insights, Refinements, and Cryptographic Alternatives}}\\
The fundamental insight behind the \ESA architecture
is that
both of the above problems can be eliminated
by collecting individuals' data
through an intermediary,
in a unified pipeline.
This intermediary (the \ESA shuffler)
explicitly manages data partitions (the \ESA crowds),
and guarantees that each partition
is sufficiently large
and of uncertain-enough size.
This is done by batching and
randomized thresholding,
which establishes differential privacy
and avoids the pitfalls of
$k$-anonymity~\cite{denning80s,crowd_blending,k-anonymity-attacks,machanavajjhala2006diversity}.
Furthermore,
this intermediary can
hide the origin of reports and
protect their anonymity---even 
when some are more detailed reports from opt-in users---and
still permit their unified analysis
(e.g., as in Blender~\cite{livshits_usenix}).
This anonymity is especially beneficial
when each individual sends more than one report,
as it can prevent their combination during analysis (cf.~\cite{apple_korolova}).

\ESA can be seen as
a refinement of existing, natural trust relationships and
best practices for
sensitive data analytics,
which assigns the responsibility for
anonymity and randomized thresholding
to an independently-trusted, standalone intermediary.
\ESA relies on cryptography to denote trust,
as well as to
strengthen protection against different attack models,
and to provide privacy guarantees
even for
unique or highly-identifying report data.
Those cryptographic mechanisms---detailed in the remainder of this paper---differ
from the cryptography typically used to protect privacy
for aggregated analysis
(or for different purposes,
like Vuvuzela or Riposte private messaging~\cite{vuvuzela,riposte}).

Other cryptography-based privacy-protection systems,
such as PDDP, Prio, and Secure Aggregation~\cite{PDDP,prio,secure_agg},
mostly share \ESA's goals
but differ greatly in their approach.
By leveraging multiparty computations,
they create virtual trusted-third-party platforms
similar to the centralized PINQ or FLEX systems,
which support
differentially-private release
of analysis results about user data~\cite{PINQ,song-elastic}.
These approaches
can improve user-data secrecy,
but must rely on added assumptions,
e.g., about clients' online availability,
clients' participation in multi-round protocols,
and attack models with an honest-but-curious central coordinator.
Also, in terms of practical adoption,
these systems 
require radical changes to engineering practice
and share RAPPOR's obstacle
of making user data overly opaque
and giving access only to statistics.

In comparison,
\ESA is compatible with existing, unchanged software engineering practices, 
since the output from the \ESA shuffler
can be gathered into databases
that have built-in guarantees of uncertainty and anonymity.
%
Furthermore,
\ESA offers
three points of control
for finding the best balance of privacy and utility,
and the best protections:
local-differential privacy, at the client,
randomized thresholding and anonymity, at the privacy intermediary,
and differentially-private release, at the point of analysis.


\begin{figure}
\begin{center}
  \includegraphics{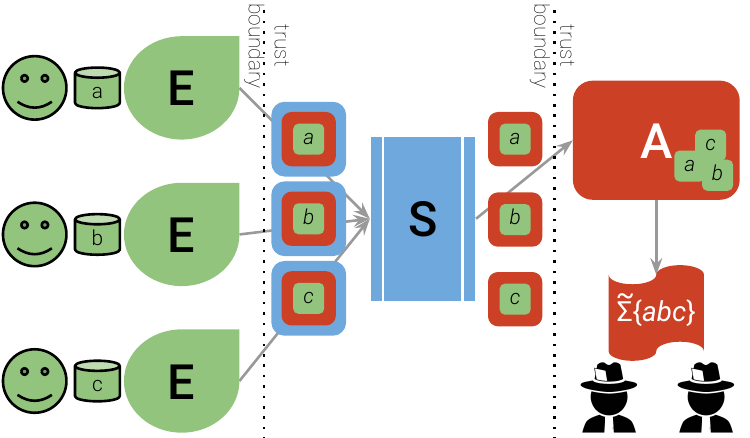}\vspace*{-2ex}
\end{center}
\caption{\label{fig:ESAArchitecture} \ESA architecture: Encode, shuffle, and analyze.}
\end{figure}

\section{The Encode-Shuffle-Analyze Architecture}
\label{sec:design-overview}

The \ESA architecture splits responsibility
between 
\emph{encoders}, \emph{shufflers}, and \emph{analyzers},
as shown in Figure~\ref{fig:ESAArchitecture}.

Encoders run on the client devices of users,
potentially as part of the software being monitored.
They
can transform 
the monitored data in a number of ways---in particular,
by converting its encoding---and
they also
use nested encryption to
guarantee which parties process it and in what order.
For privacy purposes,
encoders
control the scope, granularity, and randomness
of the data released from clients;
in particular,
they can add enough random noise
to provide users with strong guarantees,
without trust assumptions~\cite{rappor}.

Shufflers are a separate, standalone networked service
that
receives encrypted, encoded data.
They are responsible for
masking data origin and confounding data provenance
by eliminating
metadata (arrival time, order, originating IP address)
and by shuffling  (to thwart traffic analysis).
Shufflers
collect data into batches
during a lengthy interval (e.g., one day),
not only to eliminate implicit timing metadata,
but also to ensure that
every data item can get lost in a crowd
of similar items.
Large-enough batches
are forwarded for analysis
after undoing a layer of nested encryption.
Shufflers
should be
trustworthy parties
and (at least logically) completely independent
of analysis services---although their collusion is
considered in our attack model.

Analyzers are networked services
that receive shuffled data batches,
and are responsible for materializing and analyzing or releasing a database
after undoing the innermost encryption.
Their cryptographic key
determines
the specific analysis outcome
and associated privacy protection (\S\ref{sec:threat-model}).

To illustrate,
consider the app/API example
from \S\ref{sec:mot:example}
(see \S\ref{sec:evaluation} for detailed use cases).
For this example,
the \ESA architecture enables
high utility and privacy
without requiring a deluge of user data.
Since this analysis only seeks statistics about individual API uses in apps and
not multi-API patterns, \ESA encoding can transmit data as multiple,
broken-apart bitvector fragments (\S\ref{sec:design:encoder}), without necessarily
adding random noise. This fragmentation breaks apart unique patterns that might
identify users of rare applications and does not affect the utility of the
intended statistical analysis.

\ESA shuffling breaks any direct associations between data fragments of
individual users, thereby hiding their data in the crowd (\S\ref{sec:design:shuffler}). It can also add
uncertainty by randomly thresholding the data before forwarding to analysis.
By using cryptographic secret-sharing (\S\ref{sec:design:forculus}) and
blinding (\S\ref{sec:hardening:blind-crowd-id}), \ESA enables the collection of a
database of app names for analysis, without revealing any unique apps.  For
hard-to-guess, hash-based app names, this guarantee can hold even for our
strongest attack model below, where all parties, including the shuffler, collude
against users (\S\ref{sec:eval:vocab} is a concrete example).

Thanks to \ESA,
the combined data from app/API bitvector fragments
need not be considered sensitive---even if no differential privacy
was introduced by adding random noise during encoding.
Therefore, this data
is well-suited for analysis using standard  tools, techniques,
and processes.
Even so,
by adding minimal further noise at the \ESA analyzer  (\S\ref{sec:design:analyzer}),
the analysis output may be subject to
differentially-private release,
at no real loss to utility.

\subsection{Keys, Trust, and Attack Models}
\label{sec:threat-model}
Users rely on
the encoder's use of
specific keys for nested encryption
to specify
who should receive monitoring data,
what that data should contain,
and how it should be processed.
Thus,
users 
state their trust assumptions implicitly
using those keys.
This trust may be well placed,
for example,
when the cryptographic keys 
are embedded in 
software that the user must trust
(such as their operating system),
and where those keys are associated
with clear policies
for the processing of the monitored data.
The users' trust in those keys
may also be misplaced,
e.g., if the keys have been compromised,
or if different parties collude.
This provides a natural
structuring of an attack model
for the \ESA architecture:
\begin{description}
\item[Analyzer compromise]
  The analyzer is the primary attacker
  against which \ESA defends.
  Its attacks can, in particular,
  link different data received,
  and correlate it with auxiliary information---for example,
  overlapping, public databases---in
  order to breach users' privacy.
\item[Shuffler compromise]
  The shuffler is assumed to be honest but curious.
  If compromised,
  it reveals 
  the types of reports sent---when, in what frequency,
  from which IP addresses, in what data partition---but
  contents are still
  protected by the inner layer of nested encryption.
\item[Analyzer and shuffler collusion]
  If attackers
  control both the shuffler and the analyzer,
  they will see which users contribute what data---as
  if they had compromised 
  a monitoring data collection service
  not using the \ESA architecture.
  Even then,
  encoding may still provide
  some protections for users' privacy
  (e.g., as in RAPPOR~\cite{rappor}).
\item[Encoder compromise and collusion]
  In all cases, attackers may control
  most (but not all) encoders,
  and use them to send specially-crafted reports.
  In collusion with an analyzer or shuffler adversary,
  this enables more powerful attacks,
  e.g., where users' data is combined into
  a ``crowd'' comprising only Sybil data~\cite{sybil_attack}.
  Also, this enables data-pollution attacks
  on analysis, for
  which creation of a materialized database
  or a hardened encoder may be partial remedies~\cite{Abadi2004,Lie2017},
  as might be some multi-party computation schemes proposed before (e.g., Prio~\cite{prio}). However, we do not consider data pollution or Sybil attacks further in this work.
\item[Public information]
  In all cases,
  attackers are assumed to have complete access to network communications,
  as well as analysis results,
  e.g., using them to perform timing and statistical inference attacks.
\end{description}
\ESA
can still provide
differential privacy guarantees
to users under this powerful attack model.


\subsection{Encoder}
\label{sec:design:encoder}
Encoders run at users' clients,
where they mediate
upon the reporting of monitored data
and control how it is transmitted
for further processing.
As described above, 
encoders
are tied to and reflect users' trust assumptions.

Encoders locally transform and condition monitored data
to protect users' privacy---e.g., 
by coarsening or scrubbing, 
by introducing random noise,
or by breaking data items into fragments---apply
nested encryption,
and transmit the resulting reports
to shufflers, over secure channels.
Encoders also
help users' data become lost in a crowd,
by marking transmitted reports
with a \emph{crowd~ID},
upon which the \ESA shuffler
applies a minimum-cardinality constraint,
preventing analysis of any data
in crowds smaller than a threshold.

For most attack models,
encoders can provide strong privacy guarantees
even by fragmenting the users' data before transmission.
For example,
consider analysis involving
users' movie ratings,
like that evaluated in \S\ref{sec:eval:flix}.
Each particular user's set of movie ratings may be unique,
which is a privacy concern.
To protect privacy, encoders can
fragment and separately transmit users' data;
for example, 
the rating set
$\{ (m_0, r_0), (m_1, r_1), (m_2, r_2) \}$,
may be encoded
as its pairwise combinations $\langle (m_0, r_0), (m_1, r_1) \rangle$,
$\langle (m_0, r_0), (m_2, r_2) \rangle$,
and $\langle  (m_1, r_1), (m_2, r_2) \rangle$,
with each pair transmitted for independent shuffling.
Similarly, most data
that is analyzed for correlations
can be broken up into fragments,
greatly increasing users' privacy.

As an important special case,
encoders can apply randomized response to users' data, 
or similarly add random noise
in a way that guarantees local differential privacy.
While useful,
this strategy has drawbacks---as discussed earlier in \S\ref{sec:mot:lessons}.
Other \ESA encoders can offer
significantly higher utility,
while still providing differential privacy
(albeit with different trust assumptions).
For example,
\S\ref{sec:design:forculus}
discusses a \prochlo encoder
based on secret sharing.

\subsection{Shuffler}
\label{sec:design:shuffler}

The shuffler logically separates the analyzer from the clients' encoders,
introducing a layer of privacy between the two.
To help ensure that it is
truly independent, honest-but-curious, and non-colluding,
the shuffler may be run by a trusted, third-party organization, 
run by splitting its work between multiple parties,
or run using trusted hardware.
Such technical hardening is discussed in \S\ref{sec:hardening:sgx-shuffler}.

The shuffler performs four tasks: \emph{anonymization}, \emph{shuffling},
\emph{thresholding}, and \emph{batching}, described in turn.

As the first pipeline step after client encoders,
shufflers
have access to user-specific metadata about the users' reports: timestamps,
source IP addresses, routing paths, etc.
(Notably, this metadata may be
useful for admission control
against denial-of-service and data poisoning.)
A primary purpose of shufflers is to
strip all such implicit metadata,
to provide anonymity for users' reports.

However, stripping metadata may not completely 
disassociate users from reports.
In our attack model,
a compromised analyzer may monitor network traffic
and use timing or ordering information
to correlate
individual reports arriving at the shuffler with
(stripped) data forwarded to the analyzer.
Therefore,
shufflers forward
stripped data infrequently, in batches, 
and only after randomly reordering the data items.

Even stripped and shuffled data items may identify a
client due to uniqueness. For example, a long-enough API bitvector may be uniquely revealing,
and the application it concerns may be truly unique---e.g., if it was just written.
An attacker with enough auxiliary information
may be able to tie this data to a user
and thereby break privacy 
(e.g., by learning secrets about a competitor's use of APIs).
To prevent this, the shuffler can do
thresholding and throw away data items from item classes with too few
exemplars. That is the purpose of the crowd~ID, introduced into the data item by
the encoder: it allows the shuffler to count data items for each crowd~ID, and
filter out all items whose crowd~IDs have counts lower than a fixed,
per-pipeline threshold. In the API use case, the crowd~ID could be some unique application identifier, ensuring that applications with fewer than, say, 10 contributed API call vectors are discarded from the analysis.

Shuffling and thresholding have limited value for small data
sets.  The shuffler batches data items for a while (e.g., a day) or until the
batch is large enough, before processing.

\subsection{Analyzer}
\label{sec:design:analyzer}
The analyzer decrypts, stores, aggregates, and eventually retires data
received from shufflers. Although sensitivity of data records forwarded by
shufflers to the analyzer is already quite restricted, the analyzer is the
sole principal with access privileges to these records. In contrast,
in our attack model,
the analyzer's output is considered public
(although in practice, it may be subject to strict access controls).

For final protection of users' privacy, the analyzer output
may make use of 
the techniques of differentially-private release.
This may be achieved by custom analysis mechanisms
such as those in PINQ~\cite{PINQ} and FLEX~\cite{song-elastic},
or by systems such as
Airavat~\cite{Airavat}, GUPT~\cite{GUPT}, or PSI~\cite{PSI}. These ready-made
systems carefully implement a variety of differentially private mechanisms
packaged for use by an analyst without specialized expertise.

However,
it may be superfluous 
for the analyzer to produce differentially-private output.
In many cases, shuffling anonymity and encoding fragmentation may suffice
for privacy.
In other cases,
differential privacy may already be guaranteed
due to random noise at encoding
or to crowd thresholding at shuffling.
In such cases,
the database of user data
is easier to work with (e.g., using SQL or NoSQL tools),
as it requires no special protection.

For a protected database, intended to remain secret,
exceptional, break-the-glass use cases may still be supported,
even though
differentially-private release
is required
to make its analysis output public.
Such exceptional access
may be compatible with trust assumptions,
and even expected by users,
and may greatly improve utility 
e.g., by allowing the analyzer to
debug the database
and remove corrupt entries.

\subsection{Privacy Guarantees}\label{sec:dp_guarantees}
Differential privacy~\cite{DMNS,DworkRoth} is a rigorous and robust notion of
database privacy.
As described earlier, in \S\ref{sec:mot:lessons}, 
it guarantees that the distribution of analysis outputs is
insensitive to the presence or absence of a single record in the input
dataset.
Differential privacy
provides the following beneficial properties,
which may be established at three points with \ESA.


\begin{description}
\item[Robustness to auxiliary information] Differential privacy guarantees are independent of
    observers' prior knowledge, such as (possibly partial) knowledge about other
    records in the database. In fact, even if all clients but one collude with
    attackers, the remaining client retains its privacy.

\item[Preservation under post-processing] Once differential privacy is imposed
  on analysis output, it cannot be undone. This means that all
    further processing (including joint computations over outputs of other
    analysis) do not weaken the guarantees of differential privacy.

  \item[Composability and graceful degradation]
    Differential privacy guarantees 
    degrade gracefully,
    in an easily-understood manner.
    At the encoder,
    if a single user contributes their data in multiple, apparently-unrelated reports
    (e.g., because the user owns multiple devices),
    then that user's privacy only degrades linearly.
    At the analyzer,
    the privacy loss is even less---sublinear,
    in terms of $(\varepsilon, \delta)$ differential privacy---if
    a sensitive database is analyzed multiple times,
    with differentially-private release of the output.
\end{description}

Two principal approaches to practical deployment of differential privacy are \emph{local} (see earlier discussion in \S\ref{sec:mot:lessons}) and \emph{centralized}. Centralized differential privacy is typically achieved by adding carefully calibrated noise to the mechanism's outputs. In contrast with local differential privacy, the noise may be smaller than the sampling error (and thus nearly imperceptible) but this requires trusting the analyzer with secure execution of a differentially private mechanism.

The \ESA architecture offers a framework for building flexible,
private-analytics pipelines. Given an intended analysis problem, a privacy
engineer can plug in specific privacy tools at each stage of an \ESA pipeline,
some of which we have implemented in \prochlo, and achieve the desired
end-to-end privacy guarantees by composing together the properties of the individual
stages. We elaborate on such privacy tools for each stage of the \ESA pipeline,
and describe some illustrative scenarios, below.
\smallskip

\noindent\textit{\textbf{Encoder: Limited Fragments and Randomized Response}}\\
Encoding will limit the reported data
to remove superfluous and identifying aspects.
However,
even when the data itself may be uniquely identifying
(e.g., images or free-form text),
encoding can provide privacy
by fragmenting the data
into multiple, anonymous reports.
For small enough fragments
(e.g., pixel, patch, character, or n-gram),
the data may be inherently non-identifying,
or may be declared so, by fiat.

Encoders may utilize
local differential privacy mechanisms,
like  RAPPOR~\cite{rappor,apple_korolova,somesh_usenix,guha_new}.
However,
for fragments, or other small, known data domains,
users may simply probabilistically
report random values instead of true ones---a
textbook form of randomized response~\cite{Warner}.
Finally,
for domains with hard-to-guess data,
secret-share encoding 
can protect
the secrecy of any truly unique reports (see \S\ref{sec:design:forculus}).
\smallskip

\noindent\textit{\textbf{Shuffler: Anonymity and Randomized Thresholding}}\\
The shuffler's batching and anonymity guarantees
can greatly improve privacy,
e.g., by preventing traffic analysis 
as well as thwarting longitudinal analysis (cf.~\cite{apple_korolova}).

In addition,
by using the crowd~ID of each report,
the shuffler guarantees a minimum cardinality of peers (i.e., a crowd),
for reports to blend into~\cite{kanon_diffpriv,crowd_blending,Shokri-thresholding}.
Crowd~IDs may be
based on user's attributes (e.g., their preferred language)
or some aspect
of the report's payload (e.g., a hash value of its data).
If privacy concerns
are raised by revealing
crowd~IDs to the shuffler,
then they can be addressed
by using two shufflers
and cryptographic crowd~IDs blinding,
as discussed in \S\ref{sec:hardening:blind-crowd-id}.
Even na\"ive cardinality thresholding---forwarding when
$T$ reports share a crowd~ID---will improve privacy
(provably so, against the analyzer, with the secret sharing of \S\ref{sec:design:forculus}).
However thresholding
must done carefully
to avoid the 
$k$-anonymity pitfalls~\cite{denning80s,crowd_blending,k-anonymity-attacks,machanavajjhala2006diversity}.


With randomized thresholding,
the shuffler forwards only crowd~IDs
appearing more than $T+\mathit{noise}$ times, for $\mathit{noise}$
independently sampled from the normal distribution $\mathcal{N}(0,\sigma^2)$,
at an application-specific $\sigma$.
Subsuming this, shufflers can drop $d$ items from each bucket of crowd~IDs, for
$d$ sampled from a rounded normal distribution $\left\lfloor \mathcal{N}(D,
\sigma^2)\right\rceil$ truncated at $0$.
With near certainty,
the shuffler will forward
reports whose crowd~ID cardinality is much higher than $T$.

The first, similar to the privacy test in Bindschaedler 
et al.~\cite{Shokri-thresholding}, achieves differential privacy for the
crowd~ID \emph{set}.
The second provides improved privacy for the crowd~ID
counts (at the small cost of introducing a slight systematic bias of dropping
$D$ items on average for small $D$).
In combination, these two steps
address the partitioning concerns raised in \S\ref{sec:mot:lessons},
and allow the shuffler
to establish strong differential-privacy guarantees
for each crowd.

%
\smallskip

\noindent\textit{\textbf{Analyzer: Decryption and Differentially Private Release}}\\
Differentially private mechanisms at all stages of the \ESA pipeline can be deployed independently of each other; their guarantees will be complementary. 

The
analyzer, which has the most complete view of the reported data, may choose
from a wide variety of differentially private mechanisms and
data processing systems~\cite{DworkRoth,Airavat,GUPT,PSI,song-elastic}. 
While
these analysis mechanisms
will typically assume direct access to the database to be analyzed,
they do not require
the provenance of the database records---fortunately, since they are elided by the shuffler.

For an illustration of flexibility of the \ESA architecture, consider two sharply different scenarios.
In the first, users report easily-guessable items from a small, known set (e.g., common software settings).
While the data in these reports may be of low-sensitivity, and not particularly revealing,
it may have privacy implications (e.g., allowing tracking of users).
Those concerns
may be addressed by
the \ESA encoder
fragmenting users' data into multiple reports,
and by the \ESA shuffler
ensuring those reports are independent, and anonymous,
and cannot be linked together by the analyzer.

For a second example of \ESA flexibility,
consider when
users report sensitive data from a virtually unbounded domain (e.g., downloaded binaries).
By using the secret-share \ESA encoding,
or randomized thresholding at the \ESA shuffler---or both---only
commonplace, frequent-enough
data items may reach
the analyzer's database---and
the set and histogram of those items
will be differentially private.
To add further guarantees, and improve long-term privacy,
the analyzer can
easily apply differentially-private release
to their results,
before making them public.

\section{\prochlo Implementation and Hardening}
\label{sec:hardening}
Assuming separation of trust among \ESA steps, each adds more privacy to the
pipeline. Even in the worst of our threat models, the encoder's privacy
guarantees still hold (\S\ref{sec:threat-model}).
In this section, we study how \ESA can be hardened further, to make our worst threat models less likely.
Specifically, in \S\ref{sec:hardening:sgx-shuffler}, we study how
trustworthy hardware can strengthen the guarantees of \ESA,
in particular its shuffler, allowing it to be hosted
by the analysis operator. In
\S\ref{sec:design:forculus} we describe
how a novel application of secret-sharing cryptography to the \ESA encoder increases our privacy
assurances both separately and in conjunction with the shuffler. Then, in
\S\ref{sec:hardening:blind-crowd-id}, we
explore how a form of cryptographic blinding allows the shuffler to be 
distributed across distinct parties, to further protect sensitive crowd~IDs without 
compromising functionality.

\subsection{Higher Assurance by using Trustworthy Hardware}
\label{sec:hardening:sgx-shuffler}
Trustworthy hardware has been an attractive solution, especially after the
recent introduction of Intel's SGX, as a mechanism for guaranteeing correct and
confidential execution of customer workloads on cloud-provider
infrastructure.
In particular,
for the types of data center 
analytics that would be performed by the
\ESA analyzer, there have been several proposals in the recent
literature~\cite{schuster_vc3:_2015,Ohrimenko:2015:OPL:2810103.2813695,dang2017pet}.

There has been relatively less proposed for protecting the client side, e.g.,
the encoders for \ESA. Abadi articulated the vision of using secure computing
elements as client-local trusted third parties to protect user privacy~\cite{Abadi2004}.
We have also argued that trusted hardware can
\emph{validate} the user's data, to protect privacy and analysis
integrity~\cite{Lie2017}.

In this paper we focus on using trustworthy hardware to harden the \ESA Shuffler. 
In practical terms, trustworthy hardware enables the collocation
of the shuffler at the same organization hosting the analyzer, largely eliminating
the need for a distinct trusted third party; even though the same organization
operates the machinery on which the shuffler executes, trustworthy hardware
still prevents it from breaking the shuffler's guarantees
(\S\ref{sec:design:shuffler}).
This flexibility comes at a cost:
since trustworthy hardware imposes many hard constraints,
e.g. limiting the size of secure, private memory
and allowing only indirect access to systems' input/output features.
In this section,
we describe how \prochlo implements the \ESA shuffler
using Intel's SGX,
which
in current hardware realizations
provides only 92~MB of private memory
and no I/O support~\cite{sgx}.

\subsubsection{SGX Attestation for Networked Shuffler Services}
To avoid
clients having to blindly trust
the \ESA shuffler over the network,
\prochlo
employs SGX 
to assert the legitimacy of the shuffler and its associated public key.
Thereby,
per our attack models (\S\ref{sec:threat-model}),
that key's cryptographic statements
can give clients greater assurance
that their encoded messages are received by the correct, proper shuffler,
without having to fully trust
the organization hosting the shuffling service,
or its employees.

Specifically, upon startup, the shuffler generates a public/private key pair
and places the
public key in a Quote operation, which effectively attests
that ``An SGX enclave running code $X$ published public key
$\mathit{PK}_\mathit{shuffler}$.'' The shuffler's hosting organization
cannot tamper with this attestation, but can make it public
over the network.
Having fetched such an attestation, clients verify that it
(a) represents code $X$ for a known, trusted shuffler, and
(b) represents a certificate chain from Intel to a legitimate SGX CPU.
Now the client can derive an
ephemeral $K_\mathit{shuffler}$ encryption key for every data item it sends to the particular shuffler.

The shuffler must create a new key pair every time it restarts, to avoid
state-replay attacks (e.g., with a previously attested but compromised key
pair). So, unlike a shuffler running at a trusted third party, an SGX-based
shuffler will have shorter-lived public keys---say on the order of
hours. Certificate transparency mechanisms may also be used
to prevent certificate replay
by malicious operators~\cite{Maniatis:2002:SHP:647253.720277,Laurie:2014:CT:2668152.2668154}.

\subsubsection{Oblivious Shuffling within SGX Enclaves}
\label{sec:hardening:sgx-shuffler:stash-shuffling}
\emph{Oblivious shuffling} algorithms
produce random permutations of large data arrays
via a sequence of public operations on data batches---with
each batch processed in secret, in small, private memory---such
that no information about the permutation
can be gleaned by observing those operations.
Due to the limited private memory of trustworthy hardware,
random permutations of large datasets
must necessarily use
oblivious shuffling,
with the bulk of the data
residing encrypted, in untrusted external memory or disk.
Therefore,
oblivious shuffling has received significant
attention
as interest in trustworthy hardware has rekindled~\cite{Ohrimenko:2015:OPL:2810103.2813695,dinh2015m2r,dang2017pet,opaque}.

\begin{figure}
\begin{center}
  \includegraphics{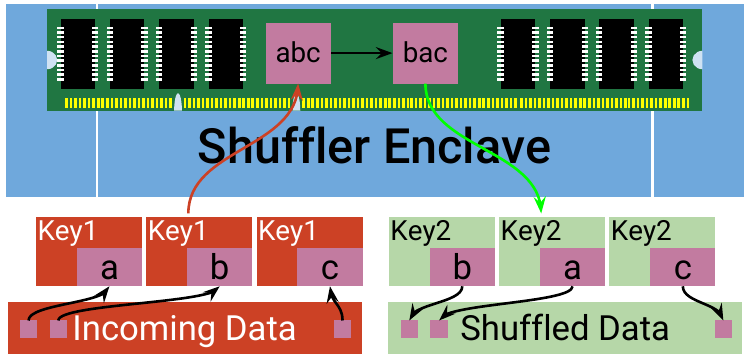}\vspace*{-2ex}
\end{center}
\caption{\label{fig:ObliviousShuffler} Primitive operation of an oblivious shuffler.\vspace*{3ex}}
\end{figure}

A primitive shuffling operation consists of reading a (small) fraction of
records encrypted with some key $\mathit{Key1}$ into private memory, randomly
shuffling those records in private memory, and then writing them back out,
re-encrypted with some other key $\mathit{Key2}$ (see
Figure~\ref{fig:ObliviousShuffler}). As long as $\mathit{Key1}$ and/or $\mathit{Key2}$
are unknown to the outside observer, the main observable signal is the sequence of
primitive shuffling operations on encrypted-record subsets.

To be secure, the shuffler must apply enough primitive shuffling operations to
its input data array so that an observer of primitive shuffling
operations gains no advantage in recovering the resulting order compared to guessing
at random; this induces the \emph{security parameter}, $\epsilon$, which is
defined as the total variation distance between the distribution of shuffled items and the
uniform distribution. Typical values range from the bare minimum of
$\epsilon = 1/N$ to a cryptographically secure $\epsilon = 2^{-128}$.

To be efficient, the shuffler must perform as few primitive shuffling operations
as possible, since each such operation imposes memory overhead (to bring a
subset into private memory, and write it back out again), cryptographic overhead
(to decrypt with $\mathit{Key1}$ and re-encrypt with $\mathit{Key2}$), and
possibly network overhead, if multiple computers are involved in running
primitive shuffling operations.

\subsubsection{State of the Art in Oblivious Shuffling}
\label{sec:oblivious-shuffling-related}
We attempted to find existing oblivious-shuffling mechanisms
that produced cryptographically-secure random permutations
with sufficient scalability and efficiency.

As described below,
we found no SGX oblivious shuffling mechanisms adequate to our purposes,
leading to the design and implementation of
our own algorithm, the Stash Shuffle (\S\ref{sec:stash-shuffle}).
In terms of scalability,
our RAPPOR experience shows
that we must handle
tens to hundreds of millions of items,
with at least 64 bytes of data and an 8-byte integer crowd~ID, each,
which in \prochlo corresponds to a 318-byte doubly-encrypted record.
%
In terms of efficiency, our driving constraint came from the
relative scarcity of SGX-enabled CPUs in data centers,
resulting from  SGX's current deployment
only as a per-socket (not per-core) feature of 
client-class CPU realizations.
We defined our metric
for efficiency
in terms of the total amount of SGX-processed data,
relative to the size of the input dataset;
thus, at 2$\times$ efficiency
SGX would read, decrypt, re-encrypt, and write out to untrusted memory
each input data item two times.

Oblivious sorting is one way to produce an oblivious shuffle: associate a
sufficiently-large random identifier (say 64 bits) with every item, by taking a
keyed hash of the item's contents, and then use an oblivious sorting algorithm
to sort by that identifier. Then the resulting shuffled (sorted) order of items
will be as random as the choice of random identifier. Oblivious sorting
algorithms are sorting algorithms that choose the items to compare and possibly
swap (inside private memory) in a data-independent fashion.

Batcher's sort is one such algorithm~\cite{batchersort}.
Its primitive 
operation reads two buckets of $b$ consecutive items
from an array of $N$ items,
sorts them by a keyed item hash, and
writes them back to their original array position.
While not restricting dataset size,
Batcher's sort requires $N/2b \times (\log_2{N/b})^2$ such
private sorting operations.
With SGX, 
$b$ can be at most 152 thousand 318-byte records.
Thus,
to apply Batcher's sort to 10 million records
(a total of 3.1 GBytes), the data processed will be 49$\times$ 
the dataset size (155 GBytes);
correspondingly,
for 100 million records (31 GBytes),
the overhead would be 100$\times$ (3.1 TBytes).
To complete this processing within a reasonable time,
at least each day,
many SGX machines
would be required to
parallelize the rounds of operations:
33 machines for 10 million records, and 330 for 100 million---a very substantial
commitment of resources for these moderately small datasets.

ColumnSort is a more efficient, data-independent sorting algorithm, 
improving on Batcher's sort's $\mathcal{O}((\log_2{}N)^2)$ rounds
and sorting datasets in just exactly 8 rounds~\cite{columnsort}.
Opaque uses SGX-based ColumnSort for private-data analytics~\cite{opaque}.
Unfortunately,
any system based on ColumnSort
has a maximum problem size
that is induced by the per-machine private-memory limit~\cite{Chaudhry2003}.
Thus,
while ColumnSort's overhead is only 8$\times$ the dataset size,
it can at
most sort 118 million 318-byte records.

Sorting is a brute-force way to shuffle: instead of producing \emph{any}
unpredictable data permutation, it picks exactly \emph{one}
unpredictable permutation and then sorts the data accordingly. A
bit further away from sorting lies the Melbourne Shuffle
algorithm~\cite{Ohrimenko2014}.  It is fast and
parallelizable, and has been applied to
privacy-data analytics in the
cloud~\cite{Ohrimenko:2015:OPL:2810103.2813695}.
Instead of picking random identifiers for items and then sorting
them obliviously, the Melbourne shuffle picks a random permutation, and then
obliviously rearranges data to that ordering.
This rearrangement uses
data-independent manipulations of the data array, without full sorting, which
reduces overhead.
Unfortunately,
this algorithm scales poorly, 
since it requires access to the entire permuted ordering in private memory.
For SGX, this means that the Melbourne Shuffle
can handle only a few dozen million items, at most,
even if we ignore storage space for actual data, computation, etc.


Finally, instead of sorting, cascade-mix networks have been proposed for
oblivious shuffling (e.g., M2R~\cite{dinh2015m2r}).
With SGX, such networks
split the input data across SGX enclaves,
partitioned to fit in SGX private memory,
and re-distribute the data after random shuffling in each enclave.
A ``cascade'' of such mixing rounds
can achieve any permutation, obliviously.
Unfortunately,
for a safe security parameter $\epsilon=2^{-64}$,
a significant number of rounds is
required~\cite{Klonowski2005}.
For our running example,
the overhead is
114$\times$ for 10 million 318-byte records,
and 87$\times$ for 100 million records.

\subsubsection{The Stash Shuffle Algorithm}
\label{sec:stash-shuffle}

Since existing oblivious shuffling algorithms
did not provide the necessary scalability,
efficiency, or randomness of the permutations,
we designed and implemented the \emph{Stash Shuffle}.
Its security analysis is in a separate report~\cite{StashShuffler}.

Our algorithm is inspired by
the Melbourne Shuffle, but does not store the permutation in
in private memory. As with other oblivious shuffle
algorithms, it reads $N$ encrypted items from untrusted memory,
manipulates them in buckets small enough to fit in private memory, and writes
them out as $N$ randomly-shuffled items, possibly in multiple
rounds. In the case of \ESA, the input consists of doubly-encrypted data items
coming from encoders, while the output consists of the inner encrypted data item
only (without crowd~IDs or the outer layer of encryption).

\begin{algorithm}[H]
\caption{The Stash Shuffle algorithm.\label{alg:StashShuffle}}
\footnotesize
\begin{algorithmic}[1]
\Procedure{StashShuffle}{Untrusted arrays $\textit{in}, \textit{out}, \textit{mid}$}
\State $\textit{stash} \gets \phi$\label{code:shuffle:distribute:1}
\For{$j \gets 0, B-1$}
\State $\textsc{DistributeBucket}(\textit{stash}, j, \textit{in}, \textit{mid})$
\EndFor
\State $\textsc{DrainStash}(\textit{stash}, B, \textit{mid})$\label{code:shuffle:drain}
\State $\textbf{FAIL}\ \textbf{on}\ \neg \textit{stash}.\textit{Empty}()$\label{code:shuffle:distribute:2}
\State $\textsc{Compress}(\textit{mid}, \textit{out})$\label{code:shuffle:compress}
\EndProcedure
\end{algorithmic}
\end{algorithm}\vspace*{-1ex}

Algorithm~\ref{alg:StashShuffle}, the Stash Shuffle, considers input ($\textit{in}$) and output ($\textit{out}$) items in $B$ sequential buckets, each holding at most $D \triangleq \lceil N/B\rceil$ items, sized to fit in private
memory. At a high level, the algorithm first chooses a random output bucket for
each input item, and then randomly shuffles each output bucket. It does that in two phases. During the
\emph{Distribution Phase} (lines~\ref{code:shuffle:distribute:1}--\ref{code:shuffle:distribute:2}), it reads in one input bucket at a time,
splits it across output buckets, and stores the split-up but as yet unshuffled, re-encrypted items in an
intermediate array ($\textit{mid}$) in untrusted memory.  During the 
\emph{Compression Phase} (line~\ref{code:shuffle:compress}), it reads the intermediate array of encrypted
items one bucket at a time, shuffles each bucket, and stores it
fully-shuffled in the output array.

The algorithm gets its name from the \emph{stash}, a private structure, whose
purpose is to reconcile
obliviousness with
the variability in item counts distributed across the output buckets.
This variability (an inherent result of balls-and-bins properties)
must be hidden from external observers,
and not reflected in non-private memory.
For this,
the algorithm caps the number traveling from an input bucket to an output bucket at
$C \triangleq D/B+\alpha\sqrt{D/B}$ for a
small constant $\alpha$.
If any input bucket
distributes more than $C$ items to an output bucket,
overflow items are instead stored in a stash---of size $S$---where
they queue, waiting to be
drained into the chosen output bucket during processing of later input buckets.

\begin{figure}[t]
\begin{center}
  \includegraphics{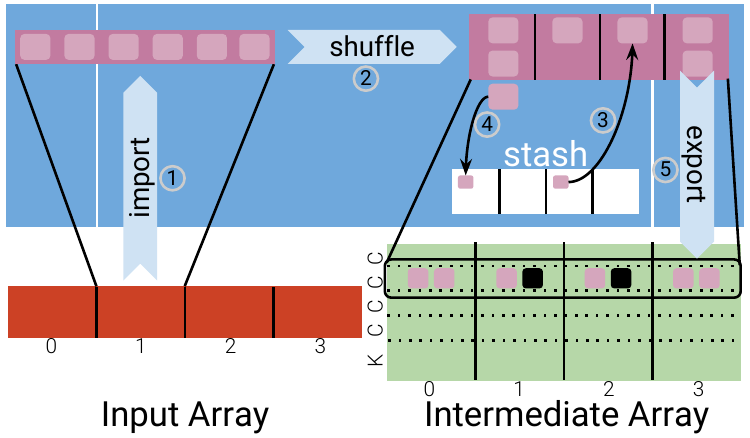}\vspace*{-2ex}
\end{center}
\caption{\label{fig:DistributionPhase} Distribution phase.}
\end{figure}

Figure~\ref{fig:DistributionPhase} illustrates the distribution phase for the
second of four buckets, where $B=4$, $D=6$, and $C=2$.  In Step 1, the bucket
is read into private memory and decrypted. It is shuffled into the 4 destination
buckets in Step 2, depositing three items in the first target bucket, one 
in the second, none in the third, and two in the last. Since the stash already
held one item for the third target bucket, that item is moved into
the third destination bucket, in Step 3. The first destination bucket got three
items but it can only deposit two, so the third item is stashed away in Step
4. Finally, the items are re-encrypted and deposited out into the intermediate
array, filling in any empty slots with encrypted dummy items to avoid leaking to the outside how items were distributed (shown black in the figure),
in Step 5.

\begin{algorithm}[H]
\caption{Distribute one input bucket.\label{alg:Distribution}}
\footnotesize
\begin{algorithmic}[1]
\Procedure{DistributeBucket}{$\textit{stash}$, $b$, Untrusted arrays $\textit{in}$, $\textit{mid}$}
\State $\textit{output} \gets \phi$
\State$\textit{targets} \gets \textsc{ShuffleToBuckets}(B, D)$\label{code:distr:targets} 
\For{$j \gets 0, B-1$}\label{code:distr:empty_stash:1}
\While{$\neg \textit{output}[j].\textit{Full}() \wedge \neg \textit{stash}[j].\textit{Empty}() $}
\State $\textit{output}[j].\textit{Push}(\textit{stash}[j].\textit{Pop}())$\label{code:distr:empty_stash:2}
\EndWhile
\EndFor
\For{$i \gets 0, D-1$}\label{code:distr:bucket:1}
\State $\textit{item} \gets \textit{Decrypt}(\textit{in}[\textit{DataIdx}(b, i)])$

\If{$\neg \textit{output}[targets[i]].\textit{Full}()$}
\State $\textit{output}[targets[i]].\textit{Push}(\textit{item})$\label{code:distr:output_chunk}
\Else
\If{$\neg \textit{stash}.\textit{Full}()$}
\State $\textit{stash}[targets[i]].\textit{Push}(\textit{item})$
\Else
\State $\textbf{FAIL}$
\EndIf
\EndIf\label{code:distr:bucket:2}

\EndFor
\For{$j \gets 0, B-1$}\label{code:distr:output:1}
\While{$\neg \textit{output}[j].\textit{Full}()$}
\State $\textit{output}[j].\textit{Push}(\textit{dummy})$
\EndWhile
\For{$i \gets 0, C-1$}
\State $\textit{mid}[\textit{MidIdx}(j, i)] \gets \textit{Encrypt}(\textit{output}[j][i])$\label{code:distr:output:2}
\EndFor
\EndFor
\EndProcedure
\end{algorithmic}
\end{algorithm}\vspace*{-1ex}

Algorithm~\ref{alg:Distribution} describes the distribution in more
detail, implementing the same logic, but reducing data copies.
$\textsc{ShuffleToBuckets}$ randomly shuffles the $D$ items of the input bucket, and $B-1$ bucket separators. The
shuffle determines which item will fall into which target bucket, stored in
$\textit{targets}$ (line~\ref{code:distr:targets}).  Then, for every output
bucket, as long as there is still room in the maximum $C$ items to output, and there are stashed away items, the output takes items from
the stash
(lines~\ref{code:distr:empty_stash:1}--\ref{code:distr:empty_stash:2}). Then the
input bucket items are read in from the outside input array, decrypted, and
deposited either in the output (if there is still room in the quota $C$ of the
target bucket), or in the stash
(lines~\ref{code:distr:bucket:1}--\ref{code:distr:bucket:2}).  Finally, if some
output chunks are still not up to the $C$ quota, they are filled with dummy
items, encrypted
and written out into the intermediate array
(lines~\ref{code:distr:output:1}--\ref{code:distr:output:2}). Note that the
stash may end up with items left over after all input buckets have been
processed, so we drain those items (padding with dummies), filling $K$ extra
items per output bucket at the end of the distribution phase
(line~\ref{code:shuffle:drain} of Algorithm~\ref{alg:StashShuffle}, which is
similar to distributing a bucket, except there is no input bucket to
distribute). $K$ is set to $S/B$, that is, the size of the stash divided by the number of buckets.

\begin{figure}[t]
\begin{center}
  \includegraphics{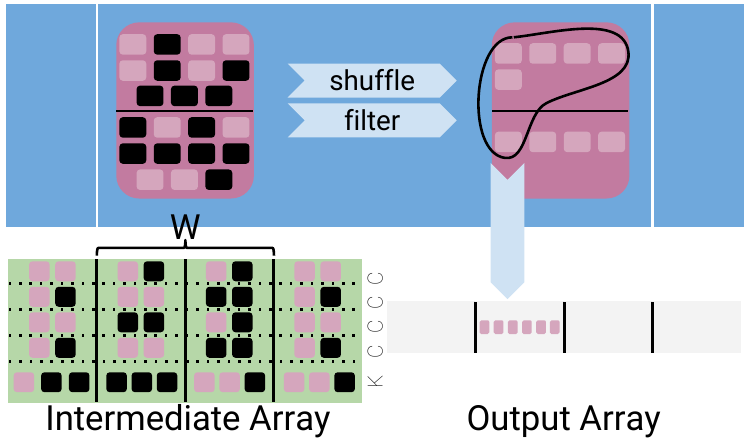}\vspace*{-2ex}
\end{center}
\caption{\label{fig:CompressionPhase} Compression phase.}
\end{figure}

\begin{algorithm}[H]
  \caption{Compress intermediate items.\\\footnotesize ($\textit{L} \triangleq \min(W, B)$ is the
  effective window size, defined to account for pathological cases where $W >
  B$.) \label{alg:Compress}}
\footnotesize
\begin{algorithmic}[1]
\Procedure{Compress}{Untrusted arrays $\textit{mid}, \textit{out}$}
\For{$b \gets 0, L-1$}
\State $\textsc{ImportIntermediate}(b, \textit{mid})$
\EndFor
\For{$b \gets L, B-1$}
\State $\textsc{DrainQueue}(b-L, \textit{mid}, \textit{out})$
\State $\textsc{ImportIntermediate}(b, \textit{mid})$
\EndFor
\For{$b \gets B-L, B-1$}
\State $\textsc{DrainQueue}(b, \textit{mid})$
\EndFor
\EndProcedure
\end{algorithmic}
\end{algorithm}\vspace*{-1ex}

Algorithm~\ref{alg:Compress} and Figure~\ref{fig:CompressionPhase}
show the compression phase.
In this phase, the intermediate items deposited by the distribution
phase must be shuffled, and dummy items must be filtered out. To do this,
without revealing information about the distribution of (real) items in output
buckets, the phase proceeds in a sliding window of $W$ buckets of intermediate
items. The window size $W$ is meant to absorb the elasticity of real item counts
in each intermediate output bucket due to the Binomial distribution. See

\begin{algorithm}
\caption{Import an intermediate bucket.\label{alg:ImportIntermediate}}
\footnotesize
\begin{algorithmic}[1]
\Procedure{ImportIntermediate}{$\textit{b}$, Untrusted array $\textit{mid}$}
\State $\textit{bucket} \gets \textit{mid}[\textit{MidIdx}(b, 0..C*B+K-1)]$\label{code:import:load}
\State $\textit{Shuffle}(bucket)$\label{code:import:shuffle}
\For{$i \gets 0, C*B+K-1$}
\State $\textit{item} \gets \textit{Decrypt}(\textit{bucket}[i])$
\If{$\neg \textit{item}.dummy$}
\State $\textit{queue}.\textit{Enqueue}(\textit{item})$
\EndIf
\EndFor
\EndProcedure
\end{algorithmic}
\end{algorithm}\vspace*{-1ex}

As Algorithm~\ref{alg:ImportIntermediate} shows, an intermediate bucket is loaded into
private memory ($C$ items per input bucket, plus another $K$ items for the final
stash drain) in line~\ref{code:import:load}, and shuffled in line~\ref{code:import:shuffle}. Then 
intermediate items are decrypted, throwing away dummies, and enqueued for export,
$D$ items at a time, into the
output array in untrusted memory. In Figure~\ref{fig:CompressionPhase}, continuing in the same example, two intermediate buckets are in private memory at steady state (i.e., $W=2$), shuffled, and streamed in their final order to the output array, $D=6$ items per output bucket, after filtering out dummies.

Even with the stash, the algorithm may fail. During distribution, the stash may
fill up or fail to fully drain in the end, and during compression the queue may
fill up when importing.  Upon failure, the algorithm aborts and starts
anew. Since intermediate items are encrypted with an ephemeral symmetric key,
failed attempts leak no information about the eventual successfully-shuffled
order.  However, failures are possible because some permutations are infeasible
to the algorithm for given parameters.  For example, the identity permutation is
infeasible unless $C\geq D$, because it would require all $D$ items of every
input bucket $b$ to land in the same $C$-sized chunk $\mathit{output}[b]$ in
Algorithm~\ref{alg:Distribution}, line~\ref{code:distr:output_chunk}.
The security analysis for a choice of $N$, $B$, $C$, $S$, and $W$
determines the security parameter $\epsilon$, which indirectly reflects
the fraction of infeasible permutations~\cite{StashShuffler}. 
Table~\ref{tab:stash-shuffle} shows a few choices for our 318-byte encrypted
items, the resulting security parameter, and the processing overhead; in the
Stash shuffle's case, we process $N$ data items and $B^2C+S$ intermediate
items. Both scaling and efficiency are significantly improved compared to prior
work (\S\ref{sec:oblivious-shuffling-related}).

\begin{table}[]
  \centering
  \small
\begin{tabular}{r|r|r|r|r||r|r}
\multicolumn{1}{c|}{$N$} &
    \multicolumn{1}{c|}{$B$} &
        \multicolumn{1}{c|}{$C$} &
            \multicolumn{1}{c|}{$W$} &
                \multicolumn{1}{c||}{$S$} &
                    \multicolumn{1}{c|}{\textbf{log($\epsilon$)}} &
                    \multicolumn{1}{c}{\textbf{Overhead}} \\ \hline
10M & 1,000 & 25 & 4 & 40,000 & -80.1 & 3.50$\times$\\ 50M & 2,000 & 30 & 4 & 86,000 & -81.8 & 3.40$\times$\\ 100M & 3,000 & 30 & 4 & 117,000 & -81.9 & 3.70$\times$\\ 200M & 4,400 & 24 & 4 & 170,000 & -64.5 & 3.32$\times$\\
\end{tabular}
\caption{Stash Shuffle parameter scenarios, their security, and relative processing overheads,
  assuming 318-byte encrypted items
  (64 data bytes and 8-byte crowd~IDs).\vspace*{1.5ex}}
\label{tab:stash-shuffle}
\end{table}

The table shows problem sizes up to 200 million items, at which point the
algorithm as we defined it above reaches its scalability limits for SGX (for
318-byte items and safe security parameters). To scale to larger problems,
the algorithm can be modified to spill internal state to untrusted
memory, increasing overall processing overhead to roughly double, or can be run
twice in succession with smaller security parameters, which has the effect of
boosting the overall security of shuffling (this is a standard technique,
presented as a default choice for the Melbourne shuffle~\cite{Ohrimenko2014}).

Oblivious shuffling benefits from parallelism.
Previous solutions, including the Melbourne shuffle,
Batcher's sort, Column sort, and cascade-mix networks consist of 4,
$(\log_2{N/b})^2$, 8, and $\mathcal{O}(\log{B})$, respectively, embarrassingly
parallel rounds, run sequentially. The Stash Shuffle 
Distribution phase parallelizes well, by replicating the stash to each worker,
and splitting a somewhat larger $K$ among workers.
However, its Compression phase is largely sequential.
Thankfully, the sequential
Compression phase is inconsequential in practice, since it involves only symmetric
cryptography, whereas the Distribution phase performs public-key operations to
derive the shared secret key $K_\mathit{shuffler}$  with which to decrypt the outer layer of
nested encryption (\S\ref{sec:eval:stash-shuffler}).

The above description is aimed at giving the reader some intuition. In our
implementation, we took care to avoid memory copies in private memory, because
they incur a cryptographic cost due to SGX's Memory Encryption Engine. For
example, we use only statically allocated private arrays organized as queues,
stacks, and linked lists, rather than standard containers employing dynamic
memory allocation; we shuffle index arrays rather than arrays of possibly
large data items; and we overlay the distribution data structures in the same
enclave memory as the compression data structures. We also took care to
minimize side
channels~\cite{branch-shadowing,stealthy-page-table-attacks,xu2015controlled};
for example, we ensure that control paths involving real and dummy items are
the same, and write the same private memory pages. In terms of performance,
our implementation is untuned. For example, to avoid issuing system OCALLs
(i.e., calls out of the enclave into untrusted space to invoke system
functionality), we memory-map our three big arrays (input batch, intermediate
array, output batch) and let virtual memory handle reading and writing from
secondary storage. Given the predictable access patterns of the stash shuffler
(sequential read of the input batch, sequential stride write of the
intermediate array, sequential read of the intermediate array, sequential
write of the output batch), we gain a benefit from intelligent prefetching
using \texttt{madvise} on the memory-mapped ranges, to reduce the cost of I/O.

\subsubsection{Crowd Cardinality Thresholding inside SGX}
Thresholding from inside an enclave with limited private memory requires some
care. It must not reveal to the hosting organization the distribution of data
items over crowd~IDs, or the values of crowd~IDs themselves; we chose to allow
the hosting organization to see the selectivity of the thresholding operation
(i.e., the number of data records that survived filtering), since the analyzer
would have access to this information anyway.

To perform thresholding, the shuffler must perform two tasks obliviously:
counting data items with the same crowd~ID, and filtering out those with crowd
IDs with counts below the threshold. In the simpler case, the crowd~ID domain---that
is, all distinct crowd~ID values---is small enough that the shuffler can
fit one counter per value in private memory; for SGX enclaves, those are crowd
ID domains of about 20 million distinct values. In that case, counting requires
one pass over the entire batch, updating counters in private memory, and another
pass to filter out (e.g., zero out) those data items with crowd~IDs with counts
below threshold. If oblivious shuffling has already taken place
(\S\ref{sec:hardening:sgx-shuffler:stash-shuffling}), these two scans of
the shuffled data items reveal no information other than the global selectivity of the
thresholding operation.
All of the problems to which we have applied \ESA have
small-enough crowd~ID domains to be countable inside private memory and can,
therefore, be thresholded in this manner.

Should a problem arise for which the crowd~ID domain is too big to fit inside private
memory, a more expensive approach is required. For example, the batch can be
obliviously sorted (e.g., via Batcher's sort) by crowd~ID, and then scanned one
more time, counting the data items in a single crowd, associating a running
count with each item (via re-encryption). In a final pass (in the opposite
scanning direction), the count of each crowd~ID is found in the first data item
scanned backwards, so all items by that crowd~ID can be filtered out. Since this
approach requires oblivious sorting anyway, it can be combined with the
shuffling operation itself, obviating the need for something more efficient like
the Stash Shuffler; in fact, this is a specialized form of more-general
oblivious relational operators, such as those proposed in Opaque~\cite{opaque}.
More efficient approaches employing approximate counting
(e.g., counting sketches) may lead to faster oblivious thresholding for large
crowd~ID domains. Nevertheless, we have yet to encounter such large crowd~ID domains in practice.

\subsection{Encoding Using Secret-Sharing Cryptography}\label{sec:design:forculus}
The \ESA architecture enables applications of novel encoding algorithms to
further protect user data.
One such encoding scheme,
constructed using \emph{secret sharing},
can guarantee
data confidentiality
for user data 
drawn from a distribution with unique and hard-to-guess items,
such as fingerprints, or random URLs.
(We call this secret sharing, instead of \emph{threshold cryptography},
so as to not overload the word ``threshold.'')
%
%
This scheme composes well with randomized thresholding at the shuffler (see \S\ref{sec:dp_guarantees}).
When combined with the blinded
crowd~IDs discussed next (in \S\ref{sec:hardening:blind-crowd-id}),
this scheme
provides strong privacy for
for both easy-to-guess, limited-domain data,
as well hard-to-guess, unique data.
These benefits are demonstrated
for a concrete example
in \S\ref{sec:eval:vocab}.

First, recall the basic idea behind Shamir secret sharing~\cite{shamir-ss}:
$t$-secret sharing splits a secret $s$
in a field $\mathbb{F}$ into arbitrary shares $s_1, s_2, \ldots$, so that any $t-1$
or fewer shares statistically hide all information about $s$.  Given any $t$
shares one can recover $s$ completely, by choosing a $(t-1)$-degree polynomial
$P \in \mathbb{F}[X]$ with random coefficients subject to $P(0) = s$. A secret
share of $s$ is the tuple $(x, P(x))$ for randomly chosen non-zero $x \in
\mathbb{F}$. Given $t$ points on the curve, Lagrange interpolation recovers
$P$ and $s = P(0)$.

From the above, a novel {\em $t$-secret share encoding} of an arbitrary string
$m$ with parameter $t$ is a pair $(c, \mathit{aux})$ constructed as follows.
Ciphertext $c$ is a deterministic encryption of the message under a
\emph{message-derived key}~\cite{mle,mle2} $k_m = H(m)$ and $\mathit{aux}$ is
a (randomized) $t$-secret share of $k_m$. It is crucial to note that these
secret shares can be computed independently by users, which enables its direct
application as an encoding scheme in the \ESA architecture.

Correctness of this encoding is easy to see: with overwhelming probability,
given $t$ independent shares  corresponding to the same ciphertext
$c$, one can decrypt it to recover $m$ by first using secret shares
$\mathit{aux}_1, \dots, \mathit{aux}_t$ to derive $k_m$ and then using $k_m$
to decrypt $c$.

This encoding protects the secrecy of $m$ with fewer instances that $t$.  With
at most $t-1$ independent shares corresponding to the same ciphertext $c$, the
values $\langle \mathit{aux}_1,\allowbreak \ldots,\allowbreak
\mathit{aux}_{t-1},\allowbreak c \rangle$ reveal no information about $m$ that
cannot anyway be guessed about $m$ by an adversary {\it a priori}. The
security analysis follows by combining statistical security of Shamir secret
sharing and by that of a message-derived key\cite{mle, mle2}. As noted before,
if $m$ comes from a distribution that is hard to guess, no information about
$m$ is leaked until a minimum of $t$ encodings is collected.

\subsection{Blinded Crowd~IDs for Shuffler Thresholding}
\label{sec:hardening:blind-crowd-id}
Using novel public-key cryptography, we enable thresholding on {\em private}
crowd~IDs in \prochlo, when a crowd~ID might itself be sensitive. Specifically,
we have designed and implemented a split shuffler, consisting of two
non-colluding parties, which together shuffle and threshold a dataset without
accessing a crowd~ID in the clear.
Rather than sending the crowd~ID encrypted to the shuffler's private key, the
encoder hashes the crowd~ID to an element of a group of a prime order $p$ as
$\mu = H(\text{crowd~ID})$. In our implementation the group is the
elliptic curve NIST P-256~\cite{nistp256} and in this presentation, we use
multiplicative notation for clarity. The encoder computes an El Gamal
encryption $(g^r, h^r \cdot \mu)$, where $(g,h)$ is Shuffler 2's public key
and $r$ is a random value in $\mathbb{Z}_p$. Shuffler 1 {\em blinds} the tuple
with a secret $\alpha \in \mathbb{Z}_p$ to compute $(g^{r\alpha}, (h^r \cdot
\mu)^\alpha)$, and then batches, shuffles, and forwards the blinded items
(with the usual accompanying data) to Shuffler~2. Shuffler~2 uses its private
key, i.e., the secret $x$ such that $h = g^x$, on input $(u, v)$ to compute
$v/u^x$ and recovers $\mu^\alpha = H(\text{crowd~ID})^\alpha$.

At the cost of three extra group exponentiations, Shuffler~2 now works
with crowd~IDs that are hashed and raised to a {\em secret} power $\alpha$.
Critically, blinding preserves the equality relation, which allows comparison
and counting.

Note that hashing alone would not have protected the crowd~ID from a
dictionary attack. With blinded crowd~IDs, Shuffler 1 cannot mount such an
attack, since it does not possess Shuffler 2's private key, and Shuffler 2
cannot mount such an attack, because it does not posses Shuffler 1's secret
$\alpha$.  As long as Shufflers 1 and 2 do not collude, this mechanism enables
them to (jointly) threshold on crowd~IDs without either party having access to
it in the clear.

This form of private thresholding is useful in several scenarios: (a) when
crowd~IDs involve ZIP code or other more personal and identifying information
that an adversary might be able to guess well; (b) when collecting data as in
\S\ref{sec:eval:flix} where low-frequency data points such as obscure movie
ratings do little to add utility but enable easy de-anonymization; and (c)
when secret-share encoding is applied to data that might be easy to guess.
\S\ref{sec:eval:vocab} presents a comprehensive use case combining blinded
crowd~IDs with secret-share encoding for analyses with sensitive crowd~IDs.

\subsection{Implementation}

The \prochlo framework is implemented in 1100 lines of C++ using OpenSSL and
gRPC~\cite{grpc,openssl}, with another 1600 lines of C++ and OpenSSL code for
cryptographic hardening. The Stash Shuffle is implemented in 2300 lines of C++
code, using SGX enclaves in ``pre-release'' mode and a combination of OpenSSL
and the Linux SGX~SDK crypto libraries~\cite{sgxsdk}. An open-source
implementation is gradually released at \url{https://github.com/google/prochlo}.
Another implementation of the \ESA architecture,
eventually intended to
include all the features of
\prochlo, is at \url{https://fuchsia.googlesource.com/cobalt}.

\section{Evaluation}
\label{sec:evaluation}
We study four data pipelines that we have ported to \prochlo. In each case, we
motivate a realistic scenario and describe encoders, shufflers, and
analyzers tailored to demonstrate the flexibility of \ESA. We seek to understand
how the utility of each analysis is affected by introducing \ESA privacy. We
use consistent parameters for the thresholding and noisy loss applied by the
shufflers (\S\ref{sec:dp_guarantees}). The thresholds are set to 20. Before
thresholding is applied, the shufflers drop $d$ items from each bucket. The
random variable $d$ is sampled from the rounded normal distribution
$\left\lfloor\mathcal{N}(D,\sigma^2)\right\rceil$, where $D=10$ and
$\sigma=2$. This guarantees $(2.25, 10^{-6})$-approximate differential privacy for
the multi-set of crowd~IDs that the analyzers receive. These settings are at
least as strong as in recent industrial deployments and in
literature~\cite{rappor,apple_korolova,abadi2016deep}.
We first evaluate our SGX-based Stash Shuffler;
the four case studies that follow use non-oblivious shufflers.

\subsection{Stash Shuffle}
\label{sec:eval:stash-shuffler}
\begin{table}[]
  \centering
\small
\begin{tabular}{r|r|r|r|r}
\multicolumn{1}{c|}{$N$} &
    \multicolumn{1}{c|}{\textbf{Distribution}} &
        \multicolumn{1}{c|}{\textbf{Compression}} &
                \multicolumn{1}{c|}{\textbf{Total}} &
                    \multicolumn{1}{c}{\textbf{SGX Mem}} \\\hline
10M & 713 s & 26 s & 738 s& 22 MB\\
50M & 3,581 s & 168 s & 1.0~h & 52 MB\\
100M & 7,172 s & 349 s & 2.1~h & 78 MB\\ 
200M & 14,267 s & 620 s & 4.1~h & 69 MB\\
\end{tabular}
\caption{Stash Shuffle execution of the scenarios in Table~\ref{tab:stash-shuffle}.
  Rows show input size;
  columns show per-phase and total execution time
  and the maximum private SGX memory used.}
\label{tab:stash-shuffle-numbers}
\end{table}

We measured the Stash Shuffler
with datasets of 318-byte items (corresponding to 64 bytes of data and 8 bytes of
crowd~ID),
on an SGX-enabled
4-core Intel i7-6700 processor, 32 GB of RAM, and a Samsung 850 1-TB SSD.
The nested
cryptography uses
authenticated encryption,
with NIST P-256 asymmetric key pairs used to derive AES-128 GCM symmetric keys.
Table~\ref{tab:stash-shuffle-numbers},
shows our measurement results,
from single-threaded runs
that form a basis for assessing scalability.
Although the two phases process roughly the same amount
of data, Distribution is far costlier, because of public-key cryptography, but parallelizes well.
Run with
10 SGX workers, approximate
execution times would be
1.6, 8.8, 17.8, and 34.1
minutes, respectively.

\subsection{Vocab: Empirical Long-tail Distributions}
\label{sec:eval:vocab}
We consider a corpus of three billion words that is representative of
English-speaking on-line discussion boards. Characteristically, the
distribution follows the power-law (Zipfian) distribution with a heavy head
and a long tail, which poses a challenge for statistical techniques such as
randomized response.  To demonstrate \prochlo's utility into recovering a
stronger signal further into the tail of the distribution, we performed the
following four experiments to privately learn word frequencies on samples of
size 10K, 100K, 1M, and 10M drawn from the same distribution. In each
experiment, we measured the number of {\em unique} words (which can be thought
of as unique candidate URLs or apps in other applications) we could recover
through our analysis.

In experiment \textsf{Crowd}, clients send \emph{unencoded} words along with a hash of
the word as the crowd~ID to a single shuffler. Against the analyzer,
this hides all words that occur infrequently and allows decoding of words
whose frequency is above the threshold.  However, a malicious shuffler may
mount a dictionary attack on the words' hashes,
and there is no privacy against
the shuffler and analyzer colluding.

Experiment \textsf{Secret-Crowd} builds on \textsf{Crowd},
but clients
encode their reported words 
using one-out-of-$t$ secret sharing,
setting $t$ to be 20, like the shuffler's crowd threshold $T$.
At a
minimal computational cost to clients (less than 50 $\mu$s per encoding),
privacy is significantly improved: uncommon words and strings drawn from
hard-to-guess data sources (such as private keys, hash values, love letters,
etc.) are private to the analyzer.
Alas, the shuffler can mount
dictionary attacks and statistical inference on crowd IDs.

Experiment \textsf{NoCrowd} uses the same
secret sharing as \textsf{Secret-Crowd},
but uses the same, fixed crowd~ID in all client reports.
This protects
against a malicious shuffler,
as it no longer can perform statistical inference or dictionary attacks
on crowd~ID word hashes.
Also, 
this  slightly improves utility by
avoiding the small noise added by the shuffler during the thresholding step.
However,
lacking a crowd to hide in,
clients now have less protection against the analyzer:
it will now receive reports
even of the most uncommonly-used words,
and can attempt
brute-force attacks
on them.

Experiment \textsf{Blinded-Crowd} offers the most compelling privacy story.
In addition to
secret-share encoding of words,
clients use blinded crowd~IDs, with two-shuffler randomized thresholding  
(\S\ref{sec:hardening:blind-crowd-id}).
Assuming no collusion,
neither the shufflers
nor the analyzer
can successfully 
perform attacks on the secret-shared words
or the blinded crowd~IDs.
Even if all parties
collude, private data from a hard-to-guess distribution (such as keys and
unique long-form text) will still be protected by the secret-share encoding.

\begin{figure}\noindent \hspace{-0.14in}
  \begin{tikzpicture}
\pgfplotsset{every tick label/.append style={font=\scriptsize},
    label style={font=\footnotesize, fill=none},
    width=1.16\columnwidth,
    height=0.9\columnwidth}
  \tikzset{
    every pin/.style={font=\footnotesize,fill=none,rectangle,rounded
    corners=3pt},
    every node/.style={font=\footnotesize},}
  \begin{loglogaxis}[
    ymin=0.2,
    xmin=3500,
    xmax=50000000,
    max space between ticks=20,
    xticklabels={, 10K, 100K, 1M, 10M},
    grid=major,
    grid style={black!15},
    legend cell align=left,
    legend style={at={(1, 0)},anchor=south east, font=\footnotesize},
    legend entries={Ground truth (no privacy),
    {\textsf{NoCrowd}~(no DP, $t\!=\!20$)},
    {{\large $\ast$-}\textsf{Crowd}
    ($\varepsilon\!=\!2\frac{1}{4},\delta\!=\!10^{-6}$)\hspace*{-1ex}},
    {\textsf{Partition}
    ($\varepsilon\!=\!2\frac{1}{4},\delta\!=\!10^{-6}$)\hspace*{-1ex}},
    {\textsf{RAPPOR} ($\varepsilon\!=\!2,\delta\!=\!0$)}
    }]
  \addplot[blue,mark=x] table {data/ground_truth.dat}
    node [pos=0.0, right=0.1cm] {4062}
    node [pos=0.34, above=0.1cm] {18665}
    node [pos=0.695, above=0.1cm] {57500}
    node [pos=1, above=0.1cm] {91260};

    \addplot[red,mark=square*] table {data/TE.dat}
    node [pos=0.0, above=0.1cm] {46}
    node [pos=0.34, above=0.1cm] {578}
    node [pos=0.695, above=0.1cm] {5921}
    node [pos=1, right=0.1cm] {28821}
    node [pos=0.05, left=0.3cm] (vertlabel) {};

    \node [anchor=south, above=0.5cm, rotate=90, style={fill=white}] at
    (vertlabel) {\# of unique words recovered};

    \addplot[cyan!10!black,mark=*] table {data/TE_plus.dat}
    node [pos=0.0, below] {32}
    node [pos=0.34, below] {371}
    node [pos=0.695, below=0.1cm] {3730}
    node [pos=1, below=0.1cm] {21972};

    \addplot[green!30!black,mark=diamond*] plot [error bars/.cd, y dir =
    both, y explicit] table[x=file_size, y=support, y
    error=error_bars]{data/RAPPOR_enhanced.dat}
    node [pos=0.34, left=0.2cm] {17}
    node [pos=0.72, left=0.1cm] {222}
    node [pos=1, right=0cm] {828};

    \addplot[purple,mark=.]
    plot [error bars/.cd, y dir = both, y explicit]
    table[x=file_size,y=support,y error=error_bars ] {data/RAPPOR.dat}
    node [pos=0.05, below=0.1cm] {2}
    node [pos=0.34, below=0.1cm] {15}
    node [pos=0.72, below=0.1cm] {122}
    node [pos=1, right=0cm] {240};
  \end{loglogaxis}
  \end{tikzpicture}
  \caption{\small \label{fig:words_experiment_chart}
    A log-log-graph
    of the number of unique words recovered (Y-axis)
    on samples of 10 thousand to 10 million Vocab words (X-axis).
    Using  results in
    The {\large $\ast$-}\textsf{Crowd} line
    results from using word hashes as the crowd-IDs,
    whereas \textsf{NoCrowd}
    offers less privacy,
    using a na\"ive threshold of 20 and no crowds.
    For comparison,
    \textsf{RAPPOR} and \textsf{Partition}
    show how
    pure local-differential privacy
    offers far less accuracy and much higher variance (error bars)
    even when
    augmented with partitions as described in \S\ref{sec:mot:lessons}.}
\end{figure}
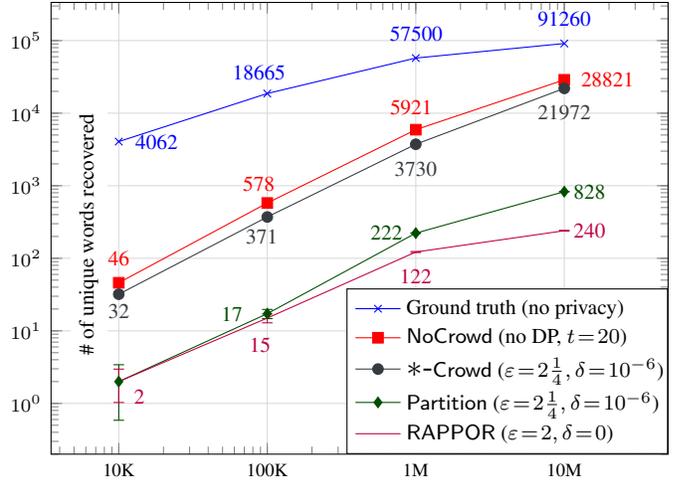

For each experiment, we compute a histogram and measure utility based on the
number of unique words recovered in the analysis. Finally, we compare \prochlo
with RAPPOR \cite{rappor} and its variant where collected reports are
partitioned by small crowd~IDs a few bits long (see the discussion of local
differential privacy in \S\ref{sec:mot:lessons}). This translates to between 4
and 256 partitions for the sample sizes in the experiment. The results are
summarized in Figure~\ref{fig:words_experiment_chart}.

Several observations are in order. 
The experiment offering the highest utility 
is \textsf{NoCrowd}
which performs no crowd-based thresholding,
but also provides 
no differential privacy guarantees,
unlike the other experiments.
Encouragingly,
the {\large $\ast$-}\textsf{Crowd} experiments
show the
utility loss due to noisy thresholding
to be very small 
compared to \textsf{NoCrowd}.
Both experiments recover
a large fraction of the ground-truth number of unique words
computed without any privacy.

The challenge of using
randomized response for long-tailed distributions is
made evident by the \textsf{RAPPOR} results,
whose utility is less than 5\% that of our \prochlo experiments.
The \textsf{Partition} results also show that
the limitations of local differential privacy 
cannot be mitigated
by following
\S\ref{sec:mot:lessons}.
and partitioning based on word hashes.
For the Vocab dataset, and the studied sample sizes,
partitioning improves RAPPOR's utility
only by between $1.13\times$ to $3.45\times$,
at the cost of relaxing guarantees from
$2$-differential privacy to $(2.25, 10^{-6})$-differential privacy.

Table~\ref{tab:words-performance} gives wall-clock running time for the Vocab
experiment across varying problem sizes. Performance was measured on an 8-core
3.5 GHz Intel Xeon E5-1650 processor with 32 GB RAM, with multiple processes
communicating locally via gRPC~\cite{grpc}. We note that these numbers
demonstrate what we naturally expect in our system design: performance scales
linearly with the number of clients and the dominating cost is public-key
crypto operations (roughly 3, 6, and 2 operations for each column,
respectively).

\begin{table}[]
  \centering
  \small
\begin{tabular}{r|r|r|r}
  \multirow{2}{*}{\textbf{\small{\# clients}}} &
  \multicolumn{2}{c|}{\textbf{\small{Encoder+Shuffler 1}}} & \multicolumn{1}{c}{\textbf{\small{Shuffler 2}}} \\
  & \small{\textsf{\{Secret-C,NoC,C\}}} & \small{{\sf Blinded-C}} &
  \small{{\sf Blinded-C}} \\ \hline
  10K &   8 s &  15 s &  7 s \\
  100K & 71 s & 153 s & 64 s \\
  1M &  713 s & 0.4 h & 643 s  \\
  10M &  2.0 h & 4.1 h & 1.8 h \\
\end{tabular}
\caption{Execution time for the Vocab experimental setup for one shuffler
  and for two shufflers (with blind thresholding).\vspace*{1.5ex}}
\label{tab:words-performance}
\end{table}

\subsection{Perms: User Actions Regarding Permissions}
\label{sec:eval:perms}
Consider monitoring new feature adoption 
in the Chrome Web browser---a rich platform with hundreds of millions of users.
For some Chrome features,
Web pages are required to ask users for access permissions;
users may respond to those requests 
by accepting, denying, dismissing, or ignoring them.
To measure feature adoption and detect Web-page abuse,
these requests
must somehow be monitored by Chrome developers.
\prochlo
provides new means
for monitoring
with both high utility and strong privacy.

From existing monitoring data,
we crafted a dataset
for the
\emph{Geolocation}, \emph{Notifications}, and \emph{Audio Capture}
permissions.
The millions of 
$\langle \text{page}, \text{feature}, \text{action bitmap}\rangle$
tuples
in the dataset
use bitmap bits
for the \emph{Grant}, \emph{Deny}, \emph{Dismiss}, and \emph{Ignore}
user actions,
since a user sometimes gives multiple responses
to a single permission prompt.
The entire dataset is privacy-relevant,
since it involves the Web pages visited,
features used,
and actions taken
by users.

In our experiments,
we performed one simple analysis of this dataset:
for each of the $3 \times 4$ feature/user-action combinations,
find the set of Web pages
that exhibited 
it
at least
100 times.
As described in Fanti et~al.~\cite{unknown_rappor},
such multidimensional analysis
over long-tail distributions (of Web pages)
is a poor fit for
local differential privacy methods.
Confirming this,
we were unable to recover more than a
few dozen Web pages, in total,
when applying RAPPOR to this dataset.

As shown in Table~\ref{sec:eval:suggest},
using \prochlo
improves utility of this simple analysis
by several orders of magnitude,
compared to the above RAPPOR baseline.
Also,
\prochlo greatly simplifies software engineering
by materializing a database of Web page names,
instead of RAPPOR's noisy statistics.
Most encouragingly,
\prochlo can
offer at least $(\varepsilon\!\!=\!\!1.2, \delta\!\!=\!\!10^{-7})$-differential privacy,
which compares very favorably.
This is achieved
by the \prochlo shuffler
using
a threshold of 100, with Gaussian noise $\sigma\!\!=\!\!4$,
and crowd~IDs
based on blinded, secret-shared
encryption of 
$\langle \text{page}, \text{feature} \rangle$.
Furthermore,
with probability $10^{-4}$,
each bitmap bit is flipped
in the encoded
$\langle \text{page}, \text{feature}, \text{action bitmap}\rangle$
tuples of \prochlo reports,
providing plausible deniability
for user actions.


\begin{table}[]
  \centering
  \small
\begin{tabular}{l|r|r|r}
  & \textbf{Geolocation} & \textbf{Notification} &
  \textbf{Audio} \\
\hline
  \textbf{Na\"ive Thresh.}
  & 6,610 & 12,200 & 620
  \\ \hline
  \textbf{Granted}
  & 5,850 & 8,870 & 440
  \\
  \textbf{Denied}
  & 5,780 & 8,930 & 430
  \\
  \textbf{Dismissed}
  & 5,860 & 9,465 & 440
  \\
  \textbf{Ignored}
  & 5,850 & 11,020 & 530 \\
\end{tabular}
\caption{Number of Web pages recovered using a na\"ive threshold
  or, for each user action, a noisy crowd threshold.\vspace*{1.5ex}}
\label{tab:eval:perms}
\end{table}

\subsection{Suggest: Predicting the Next Content Viewed}
\label{sec:eval:suggest}
Next we consider a use case
that concerns highly privacy-sensitive input data
and existing, hard-to-change analysis software, in the context of YouTube.
For this use case,
the analysis
is custom code for training 
a multi-layer, fully-connected neural network
that predicts
videos that users may want to view next,
given their recent view history.
Content popularity follows a long-tail distribution,
and the input data
consists of longitudinal video views of different users.
The resulting state-of-the-art deep-learning sequence model
is
based on ordered views
of the
half-million most popular videos,
each with at least tens-of-thousands of views.

Similar sequence prediction models
are common, and can be widely useful
(e.g., for predicting what data to load into caches
based on users' software activity).
However,
sequence-prediction analysis
is inherently at odds with privacy:
its input is a longitudinal history of individual users' data---and
some parts of that history
may be damaging or embarrassing to the user,
while other parts may be highly identifying.
In particular, for the video view input data,
any non-trivial sequence of $n$ views
is likely close to unique,
as it forms
an ordered $n$-tuple over 
500,000-item domain.
%

For this example use case,
it is neither possible to modify existing, complex analysis code 
nor to remove the inherently identifying sequence ordering
upon which the analysis relies.
Therefore,
to protect users' privacy,
we implemented a \prochlo encoding step
that fragmented each user's view history into short, disjoint $m$-tuples ($m \ll n$),
and relied on \prochlo shuffling
to guarantee that
only such tuples could be analyzed---i.e., only
anonymous, disassociated 
very short
sequences of views of very popular videos.

This construction is appealingly simple
and provides a concrete, intuitive privacy guarantee:
for small-enough $m$
(i.e., as $m$ approaches 1),
any single $m$-tuple output by the encoder
can be identifying or damaging, but not both.
Thus, privacy can be well protected
by the shuffler
preventing
associations from being made between any two disjoint tuples.

Fortunately,
because recent history is the best predictor of future views,
a model trained
even with $3$-tuples correctly predicts the next view more than 1 out of 8 times,
with around 90\% of the accuracy of a model trained
without privacy.
Training each model to convergence
on about 200 million longitudinal view histories
takes two days on a small cluster with
a dozen NVIDIA Tesla K20 GPUs,
using TensorFlow~\cite{tensorflow}.
Even more relevant is that
this privacy-preserving model
has equivalent quality
as the best-known YouTube model for this task from about a year ago---when
evaluated using an end-to-end metric
that models presenting users with multiple suggestions for what to view next.

\subsection{Flix: Collaborative Filtering}
\label{sec:eval:flix}
We next consider a prediction task without the strong locality
inherent to
our good results for next-viewed content
in the previous section.
This task is to infer users' ratings for content
given the \emph{set} of each user's content ratings.
Work on this problem
was supercharged by the \$1M Netflix Prize challenge
for improving their proprietary recommender system.
Famously,
Narayanan and
Shmatikov~\cite{narayanan2008robust}
managed to deanonymize users in the challenge's
large corpus of training data
by exploiting
auxiliary data
and 
the linkability of users' movie preferences.
Their successful attack led to the
cancellation of a followup competition.

We demonstrate that the \ESA architecture permits collection of data that
simultaneously satisfies dual (and dueling) objectives of privacy and utility
for collaborative filtering. As in the previous example, sending a complete
vector of movie ratings exposes users to linking attacks. The equally
unsatisfying alternative is to guarantee privacy to all users by randomizing
their vectors client-side. Since the ratings and the movies are sensitive
information, both need to be randomized, effectively destroying data utility.

To enable utility- and privacy-preserving collection of data we identify
sufficient statistics that can be assembled from anonymized
messages drawn from a small domain. We observe that many of 
the most relevant analysis methods of collaborative filtering
comprise two distinct steps: (i) computing the \emph{covariance matrix}
capturing item-to-item interactions, and
(ii) processing of that matrix,
e.g., for factorization or de-noising.
Computation on the sensitive data of users' movie ratings is 
performed only in the first step,
which can be supported by the \ESA architecture, as described below.

Let the rating given to item $i$ by user $u$ be $r_{ui}$, the set of items rated
by user $u$ be $I(u)$ and the set of users who rated item $i$ be $U(i)$. Towards
the goal of evaluating the covariance matrix we compute two intermediate
item-by-item matrices $S$ and $A$ defined as $S_{ij}=|U(i)\cap U(j)|$ and
$A_{ij}=\sum_{u\in U(i)\cap U(j)}r_{ui}r_{uj}$. An approximation to the
covariance matrix is given by $(A_{ij}/S_{ij})$.

We describe how $A$ is computed ($S$ is treated analogously). By pivoting to
users, we represent $A$ as follows: $A_{ij} =
\sum_u\sum_{i,j\in I(u)}r_{ui}r_{uj}$. Thus, it is sufficient for each user to
send its contribution to $A$ that consists of all $(i,r_{ui},j,r_{uj})$
four-tuples where $i,j\in I(u)$ (by symmetry only tuples where $i\leq j$ are
needed).

Even though most four-tuples are unlikely to lead to re-identification, a truly
unique item-rating four-tuple could allow linking all of the items of the
contributing user. To minimize this possibility we pursue three complementary
approaches. First, only a random set of four-tuples is sent by each user, capped
in cardinality. Second, users replace a fixed fraction (10\% in our experiments)
of the movie identifiers in their reports with a randomly sampled one (this alone affords 
$2.2$-differential privacy for the set of rated movies).
Third, each four-tuple $(i,r_{ui},j,r_{uj})$ is tagged with two
crowd~IDs, one for $(i,r_{ui})$ and one for $(j,r_{uj})$, adding a layer of
nested encryption and a second shuffler to the pipeline. This way, each
item-rating combination that reaches the analysis server appears more than a
threshold number of times.

\begin{table}[]
  \centering
  \small
\begin{tabular}{r|r|r|r|r}
\multirow{2}{*}{\textbf{\small{\# Movies}}}
  & \multirow{2}{*}{\textbf{\small{\# Users}}}
  & \multirow{2}{*}{\textbf{\small{\# Reports}}}
  & \multicolumn{2}{c}{\textbf{Score (RMSE)}} \\
&  &  & no privacy & \prochlo \\ \hline
200 & 90K & 1.77M & 0.9579 & 0.9595$^\dagger$ \\
2K & 353K & 335M &  0.9414 & 0.9420\hphantom{$^\dagger$} \\
18K & 480K & 22.6B &  0.9222 & 0.9242\hphantom{$^\dagger$} \\
\end{tabular}
\caption{Utility of the Flix evaluation; lower numbers are better. ($\dagger$To account for sparsity, the threshold was set to 5.)\vspace*{1.5ex}}
\label{tab:flix:utility}
\end{table}

We perform our experiments on a dataset whose characteristics precisely match
that of the Netflix Prize dataset: the number of users is 480K, the number of
movies is 18K, the ratings are integers between 1 and 5. Utility is measured as
the root mean square error (RMSE) and reported relative to the same benchmark
used as part of the competition. Two smaller datasets (200 and 2{,}000 movies)
are selected randomly from the main set. As seen in
Table~\ref{tab:flix:utility}, the RMSE with and without \prochlo privacy is
similar across datasets.


\section{Conclusions}
\label{sec:conclusion}
Although a long-standing issue,
the privacy of users' software monitoring data has
recently become a pressing concern.
Fortunately,
those concerns can be addressed
in a manner that
simultaneously 
permits high-utility analysis,
is compatible with standard software engineering practice,
and provides users with strong privacy guarantees.

This paper has described how
to address those privacy concerns
in the context of the \ESA architecture,
and 
its \prochlo implementation.
To offer good means of balancing privacy and utility,
and to minimize trust,
\prochlo
introduces both
new cryptographic primitives
and a new algorithm for oblivious shuffling,
and also
relies on the advanced technologies
of trusted computing and differential privacy.
Even so,
\prochlo remains a
relatively simple, easy-to-understand system,
and a straightforward realization of the \ESA architecture.
However,
as a framework for balancing privacy and utility,
\ESA is flexible enough
to permit many implementations,
and the use of the most appropriate techniques
for different data-collection and analysis scenarios.

\acks This paper is dedicated to the memory of Andrea Bittau, our colleague
who wrote much of \prochlo. We thank Kunal Talwar for his help analyzing
Stash Shuffle's security properties.
We thank the anonymous reviewers for their detailed feedback,
and Mart\'{i}n Abadi, Johannes Gehrke, Lea
Kissner, No\'{e} Lutz, and Nicolas Papernot for their valuable advice
on earlier drafts.
Our shepherd, Nickolai Zeldovich, provided invaluable help with
this final paper version.

\clearpage
\bibliographystyle{acm}
\let\OLDthebibliography\thebibliography
\renewcommand\thebibliography[1]{
  \OLDthebibliography{#1}
  \setlength{\parskip}{0.5ex}
  \setlength{\itemsep}{0.2ex plus 0.3ex}
}

\end{document}